\documentclass[aps,prd,onecolumn,nofootinbib,notitlepage,showkeys,superscriptaddress,preprintnumbers]{revtex4-2}

\usepackage{graphicx}
\usepackage{dcolumn}
\usepackage{bm}
\usepackage{subcaption}
\usepackage{tikz-feynman}
\usepackage{amsmath,amssymb}
\usepackage{float}
\usepackage{multirow}
\usepackage{slashed}
\usepackage{xcolor}
\usepackage{physics}
\usepackage{multirow}
\usepackage[colorlinks=true, pdfstartview=FitV, bookmarks=true, bookmarksnumbered=true, breaklinks]{hyperref}

\usepackage{mathtools,braket}
\usepackage{soul}
\usepackage{lipsum}  
\usepackage{color}
\definecolor{blue}{rgb}{0.0, 0.0, 1.0}
\definecolor{red}{rgb}{1.0, 0.0, 0.0}
\definecolor{royalblue}{rgb}{0.0, 0.14, 0.4}
\hypersetup{linkcolor=royalblue, citecolor=blue, urlcolor=royalblue}

\usepackage{hyperref}
\hypersetup{colorlinks=true,citecolor=blue,linkcolor=blue,urlcolor=blue}

\usepackage[mathlines]{lineno}
\usepackage{amssymb}
\usepackage{amsmath}
\usepackage{slashed}
\usepackage{xcolor}
\usepackage{graphicx}
\usepackage[normalem]{ulem}
\usepackage{color}

\usepackage{tikz,xcolor,hyperref}
\definecolor{lime}{HTML}{A6CE39}
\DeclareRobustCommand{\orcidicon}{%
	\begin{tikzpicture}
	\draw[lime, fill=lime] (0,0) 
	circle [radius=0.16] 
	node[white] {{\fontfamily{qag}\selectfont \tiny ID}};
	\draw[white, fill=white] (-0.0625,0.095) 
	circle [radius=0.007];
	\end{tikzpicture}
	\hspace{-2mm}
}

\foreach \x in {A, ..., Z}{%
	\expandafter\xdef\csname orcid\x\endcsname{\noexpand\href{https://orcid.org/\csname orcidauthor\x\endcsname}{\noexpand\orcidicon}}
}

\allowdisplaybreaks

\begin{document}
\title{A general formalism for electromagnetic production of two mesons on the nucleon}

\author{Reddy Malik Priandika} 
\email[E-mail: ]{reddympriandika@gmail.com}
\affiliation{Departemen Fisika, FMIPA, Universitas Indonesia, Depok 16424, Indonesia}
\author{Samson Clymton\orcidB{}}
\email[E-mail: ]{samson.clymton@apctp.org} %
\affiliation{Asia Pacific Center for Theoretical Physics (APCTP), Pohang, Gyeongbuk 37673, Republic of Korea}
\author{Terry Mart\orcidC{}}
\email[E-mail: ]{terry.mart@sci.ui.ac.id}
\affiliation{Departemen Fisika, FMIPA, Universitas Indonesia, Depok 16424, Indonesia}
\date{\today}

\begin{abstract}
We develop a general formalism for the electromagnetic production of two mesons on the nucleon, $\gamma_{\rm v}N\to\Phi_1\Phi_2B$, that can be used in single-channel as well as coupled-channel analyses over a wide range of energies. In the spirit of the decomposition introduced by Berends \textit{et al.} for single-meson electroproduction, the hadronic current is expanded in a set of 16 gauge- and Lorentz-invariant matrices, four of which 
vanish for real photons. The corresponding Lorentz-invariant (Dirac) amplitudes can be obtained from any reaction model and are mapped onto 16 CGLN-like Pauli amplitudes, from which the five-fold differential cross section and its transverse, longitudinal, and interference response functions follow directly. Although these matrices are not all linearly independent, they follow directly from the covariant decomposition of the current, so the amplitudes can be read off from the Feynman diagrams and the observables take a simpler, more symmetric form. As a first application, we calculate the $\gamma p\to\pi^0\pi^0p$ and $\gamma p\to\pi^+\pi^0n$ reactions near threshold from tree-level nucleon Born terms. This example shows explicitly how the Dirac amplitudes are read off from Feynman diagrams and how gauge invariance, which is violated by the individual diagrams, is restored in the total amplitude and preserved when hadronic form factors are included. Compared with the MAMI data, the Born terms overestimate the $\pi^0\pi^0p$ total cross section, which can be reproduced up to $E_\gamma\approx 370$~MeV with a soft form factor, whereas they account for only about 40\% of the $\pi^+\pi^0n$ cross section at $E_\gamma=400$~MeV. Moreover, the calculated $\pi^0\pi^0$ invariant-mass spectrum peaks too close to threshold, independently of the form factor. These deficiencies point to the $\Delta(1232)$, $\rho$-meson, and isoscalar $\pi\pi$ final-state mechanisms, all of which could be incorporated in the present framework in future work. The formalism thus provides a versatile basis for future theoretical and phenomenological studies of double-meson photo- and electroproduction.
\end{abstract}
\maketitle

\section{Introduction} \label{sec:intro}
The electromagnetic production of two mesons on the nucleon is one of the principal tools for investigating the spectrum of baryons and the dynamics of meson production. In contrast to single-meson production, the final state consists of three particles, and its kinematics is fully determined by five independent variables at fixed total c.m. energy $W$ and photon virtuality $\mathcal{Q}^2$. Besides total and differential cross sections, one can therefore measure the invariant-mass distributions of the meson-baryon and meson-meson subsystems, angular correlations between the final-state particles, and Dalitz-plot densities. These observables make it possible to reconstruct intermediate resonances, to identify sequential decay chains, and to separate competing reaction mechanisms, and they consequently constrain theoretical models much more strongly than the observables of two-body final states \cite{Roberts:2004mn,Kamano:2008gr,Kamano:2009im,Arenhovel:2014qoj}. The price to pay is a considerably larger number of polarization observables: a complete experiment for the photoproduction of two pseudoscalar mesons requires not only single- and double- but also triple-polarization observables \cite{Roberts:2004mn,Arenhovel:2014qoj}.

The additional kinematic degrees of freedom also provide enhanced sensitivity to the interplay between resonant and nonresonant contributions, and thus help to clarify the role of background mechanisms that are difficult to isolate in single-meson production \cite{Kamano:2009im,Mart:2006dk}. Baryon resonances manifest themselves as structures in the meson-baryon invariant masses, whereas meson resonances and meson-meson final-state interactions show up in the meson-meson subsystem. Nonresonant mechanisms, such as Born terms, meson-exchange and contact (Kroll-Ruderman-type) processes, and final-state rescattering, populate the multidimensional phase space in characteristically different ways \cite{Kamano:2009im,Fix:2005if}. A simultaneous analysis of invariant-mass spectra, angular distributions, and polarization observables therefore imposes stringent constraints on any theoretical description of the reaction \cite{Roberts:2004mn,Mai:2023cbp,Wang:2024byt}.

These features make double-meson production an indispensable tool for baryon spectroscopy. Many higher-lying nucleon and $\Delta$ resonances are expected to decay predominantly into three-body final states through quasi-two-body channels such as $\pi\Delta$, $\rho N$, and $\sigma N$, or through cascades via lower-lying resonances, $N^*\to\pi N^{*\prime}\to\pi\pi N$ \cite{Burkert:2017djo,CBELSATAPS:2015tyg}. Since these channels end up in double-meson final states, double-meson photo- and electroproduction provide direct access to the dominant decay modes of such states and to their electromagnetic transition amplitudes. In particular, the $\pi^+\pi^-p$ photo- and electroproduction data from CLAS have become a major source of information on the photocouplings of nucleon resonances and on the $Q^2$ evolution of their electrocouplings \cite{Aznauryan:2005tp,CLAS:2018drk,CLAS:2026qhw,Mokeev:2026plj}. Double-meson production has therefore become an essential ingredient of modern dynamical coupled-channel and multichannel partial-wave analyses, enabling a more reliable extraction of resonance parameters and the search for missing nucleon resonances \cite{Kamano:2009im,Kamano:2013iva}.

From the theoretical point of view, double-meson production is considerably more involved than single-meson production. Besides the multidimensional kinematics of the three-body final state, the amplitude receives contributions from a much richer set of topologically distinct mechanisms, including two independent $t$-channel exchanges associated with the two outgoing mesons, as well as $s$- and $u$-channel processes in which the photon couples to different baryon lines. Already at tree level, individual diagrams are not gauge invariant, and the introduction of hadronic form factors, which is unavoidable in a phenomenological description, requires special care to maintain current conservation \cite{Haberzettl:1997jg,Haberzettl:1998aqi,Davidson:2001rk}. A field-theoretical framework in which gauge invariance is preserved order by order in the number of dressed loops has been formulated by Haberzettl, Nakayama, and Oh \cite{Haberzettl:2018hcl}. Furthermore, the nonresonant part of the amplitude is much more important and more model dependent than in single-meson production; for instance, near-threshold predictions for the $\pi^0\pi^0$ channel depend strongly on whether pseudoscalar or pseudovector $\pi NN$ couplings are used \cite{Hirata:2002tp,Fix:2005if}. A general framework should therefore treat all of these mechanisms on an equal footing, preserve gauge invariance, and remain applicable to both photo- and electroproduction \cite{Roberts:2004mn,Haberzettl:2018hcl}.

Experimental and theoretical studies of double-meson electromagnetic production on the nucleon date back to the 1960s. Pioneering measurements of double-pion photoproduction were reported in the 1960s and early 1970s \cite{Chasan:1960zz,Allaby:1966,Piazza:1970tr,Beneventano:1973gx}, and one of the first theoretical descriptions of the $\gamma p\to\pi^+\pi^-p$ reaction was given by Pittner and Urban~\cite{Pittner:1972yx}, who employed an effective Lagrangian model. Since then, both experimental~\cite{Kusumegi:1972jb,Aachen-Hamburg-Heidelberg-Muenchen:1973oxa,Braghieri:1994rf,Kotulla:2003cx,GDH:2003fjc,Aznauryan:2005tp,GDH:2005jgl,CrystalBallatMAMI:2009iym,Zehr:2012tj,CBELSATAPS:2015kka,He:2016lwg,CLAS:2018drk,Dieterle:2020vug,TAPS:2022ijf,A2:2022ipx,A2:2023ckm,CLAS:2023mfc,CLAS:2026auu,CLAS:2026qhw,CLAS:2026ywm,Mokeev:2026plj} and theoretical~\cite{Dosch:1968mv,Bhargava:1968nza,Huang:1968kcy,Luke:1968mxp,Benmerrouche:1994ue,Bernard:1994ds,GomezTejedor:1995pe,Fix:2005if,Dusaev:2017qfp,Kang:2024fsf} studies have progressed substantially. The threshold region has proven particularly instructive. Chiral perturbation theory (ChPT) predicted that the $\gamma p\to\pi^0\pi^0p$ cross section near threshold is strongly enhanced by pion loops, because the tree-level contributions to this channel are suppressed \cite{Bernard:1994ds,Bernard:1996ft}, and this prediction was confirmed by precise measurements at MAMI \cite{Kotulla:2003cx,Zehr:2012tj}. Effective-Lagrangian models, in which the $\pi^+\pi^-p$ channel at low energies is dominated by the $\Delta$ Kroll-Ruderman and pion-pole mechanisms, describe the total cross sections of the charged-pion channels reasonably well, but their predictions for $\pi^0\pi^0p$ near threshold differ considerably: some models underestimate the data unless $\pi\pi$ rescattering or pion loops are included \cite{Roca:2002vd,Fix:2005if}, while others reproduce them, depending on the $\pi NN$ coupling employed \cite{GomezTejedor:1995pe,Hirata:2002tp}. More recently, a covariant ChPT calculation with explicit $\Delta(1232)$, $N(1440)$, and $\rho(770)$ degrees of freedom described the $\gamma p\to\pi^+\pi^0n$ data, whereas a satisfactory description of $\gamma p\to\pi^0\pi^0p$ required the inclusion of the isoscalar $S$-wave $\pi\pi$ final-state interaction \cite{Kang:2024fsf}.

Despite this progress, a general decomposition of the double-meson production amplitude comparable to that introduced by Berends \textit{et al.}~\cite{Berends:1967vi} for single-meson production has, to the best of our knowledge, not been developed. In their pioneering work, the single-meson amplitude was decomposed into a complete set of Lorentz- and gauge-invariant structures, and the cross section and polarization observables were expressed in a compact and systematic form in terms of the corresponding invariant amplitudes or, equivalently, the CGLN amplitudes \cite{Chew:1957tf,Berends:1967vi}. This framework has since become the standard tool for theoretical and experimental studies of pion, eta, and kaon photo- and electroproduction~\cite{Drechsel:1992pn,Knochlein:1995qz,Drechsel:2007if,Mart:2026tyg}. For double-meson production, the general structure of the observables has been worked out in the helicity basis \cite{Roberts:2004mn,Arenhovel:2014qoj}, and the gauge-invariant field-theoretical structure of the current has been clarified \cite{Haberzettl:2018hcl}. What is still missing is a covariant decomposition of the current into gauge-invariant structures, together with a prescription that converts the associated invariant amplitudes into cross sections. Such a decomposition would allow the amplitudes of any reaction model to be inserted directly, irrespective of whether they originate from tree-level diagrams, unitarized models, or coupled-channel calculations, and would make it straightforward to include or exclude individual baryon and meson resonances, as is routinely done in single-meson production. It would also provide a natural starting point for multipole analyses and for phenomenological applications analogous to those carried out for single-meson production~\cite{Mart:1996ay,Miyagawa:2006kj,Mart:2004at,Mart:2011ey,Mart:2013fia,Mart:2019mtq,Clymton:2021wof,Mart:2026tyg}.

Motivated by these considerations, we develop a general formalism for the electromagnetic production of two mesons on the nucleon. Following the approach established for single-meson production, the hadronic current is decomposed into a set of 16 gauge-invariant matrices, whose scalar coefficients, the Dirac amplitudes, can be determined within any given reaction model. These matrices are subsequently reduced to Pauli operators, yielding a set of CGLN-like amplitudes in terms of which the five-fold differential cross section and its transverse, longitudinal, and interference response functions are expressed. To demonstrate the use of the formalism, we apply it to the reactions $\gamma p\to\pi^0\pi^0p$ and $\gamma p\to\pi^+\pi^0n$ near threshold, using only the nucleon Born terms. This deliberately minimal example shows how the Dirac amplitudes are read off from Feynman diagrams, how gauge invariance is restored in the total amplitude and maintained in the presence of hadronic form factors, and how the comparison with the total cross sections and invariant-mass distributions measured at MAMI \cite{Zehr:2012tj} reveals which mechanisms are missing.

This paper is organized as follows. Section~\ref{sec:formalism} presents the formalism for double-meson photo- and electroproduction, including the kinematics, the hadronic current, the 16 Lorentz- and gauge-invariant matrices together with their Lorentz-invariant scalar amplitudes, the reduction of the Dirac amplitudes to Pauli amplitudes, and the formulation of the cross section. Section~\ref{sec:example} illustrates the application of the formalism to the double-pion photoproduction processes $\gamma p\to\pi^0\pi^0p$ (Sec.~\ref{sec:res_pi0_pi0}) and $\gamma p\to\pi^+\pi^0n$ (Sec.~\ref{sec:res_pip_pi0}), and the numerical results are presented and discussed in Sec.~\ref{sec:numerical}. Section~\ref{sec:summary_and_conclusion} summarizes our findings. The relations between the Pauli and Dirac amplitudes, as well as the components of the hadronic tensor, are given in the Appendices.

\section{Theoretical Formalism}
\label{sec:formalism}
\subsection{Kinematics}
\label{subsec:kinematics}
\begin{figure}[ht]
\includegraphics[width=0.4\columnwidth]{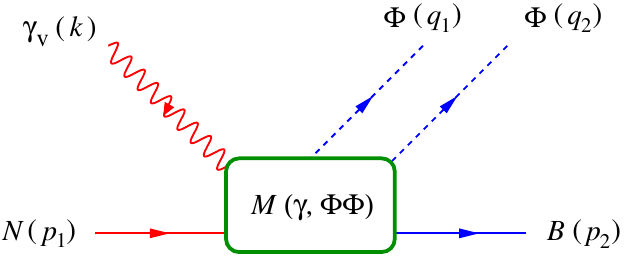} 
\caption{\label{fig:feynman_general} Schematic diagram for double-meson virtual photoproduction on the nucleon. Particles in the initial and final states are indicated with the red and blue colors, respectively.} 
\end{figure}

The electromagnetic production of two mesons is an inelastic scattering process. In this process, an interaction between an electron $e$ and nucleon $N$ produces a final state consisting of scattered electron $e'$, a pair of mesons $\Phi_1$ and $\Phi_2$, and a baryon $B$, and can be expressed as
\begin{align}
    e (k_1) + N(p_1) \to e'(k_2) + \Phi_1 (q_1) + \Phi_2 (q_2) + B(p_2)
\end{align}
where $k_1$, $k_2$, $p_1$, $p_2$, $q_1$, and $q_2$ denote the four-momentum of the incoming electron, scattered electron, nucleon, baryon, first and second meson, respectively.  This process can also be written in terms of virtual photoproduction, 
\begin{align}
\label{eq:virtual_photo}
    \gamma_\text{v} (k) + N(p_1) \to  \Phi_1 (q_1) + \Phi_2 (q_2) + B(p_2)
\end{align}
where $k = k_1 - k_2$ is the virtual photon momentum. The corresponding process is diagrammatically illustrated in Fig.~\ref{fig:feynman_general}.

From the five four-momenta involved in the double-meson virtual photoproduction process given in Eq.~\eqref{eq:virtual_photo}, one can construct 10 Mandelstam invariants, i.e.,
\begin{align}
    s      &= (p_1+k)^2,    & t      &= (p_1-p_2)^2, & u      &= (p_2-k)^2,  & s_{12} &= (q_1+q_2)^2,  & s_{13} &= (q_1+p_2)^2, \nonumber\\
    s_{23} &= (q_2+p_2)^2,  &t_1    &= (q_1-k)^2,    & t_2    &= (q_2-k)^2,   & u_1    &= (p_1-q_1)^2, & u_2 &= (p_1-q_2)^2,
\end{align}
where $m_1$ and $m_2$ denote the masses of the initial- and final-state baryons, respectively, while $\mu_1$ and $\mu_2$ are the masses of the two produced mesons. The four-momentum conservation relation,
\begin{align}
    p_1 + k = p_2 + q_1 + q_2,
\end{align}
provides five independent constraints among these 10 invariants. These constraints are obtained by successively isolating each of the five four-momenta in the conservation equation and squaring the resulting expressions, yielding
\begin{align}
    s + u + t_1 + t_2           &= 2k^2 + m_1^2  + m_2^2  + \mu_1^2 + \mu_2^2, \nonumber\\
    s + t + u_1 + u_2           &= k^2  + 2m_1^2 + m_2^2  + \mu_1^2 + \mu_2^2, \nonumber\\
    s_{13} + s_{23} + t + u     &= k^2  + m_1^2  + 2m_2^2 + \mu_1^2 + \mu_2^2, \\
    s_{12} + s_{13} + t_1 + u_1 &= k^2  + m_1^2  + m_2^2  + 2\mu_1^2 + \mu_2^2, \nonumber\\
    s_{12} + s_{23} + t_2 + u_2 &= k^2  + m_1^2  + m_2^2  + \mu_1^2 + 2\mu_2^2.\nonumber
\end{align}
Summing the five equations, in which each Mandelstam invariant appears exactly twice, yields the compact identity
\begin{align}
    s + s_{12} + s_{13} + s_{23} + t + t_1 + t_2 + u + u_1 + u_2
    = 3\!\left(k^2 + m_1^2 + m_2^2 + \mu_1^2 + \mu_2^2\right).
\end{align}
The five independent constraints reduce the number of independent kinematic
variables from 10 to 5. We choose three Lorentz-invariant variables with clear
physical significance: the total center-of-mass energy squared, $s$, the
di-meson invariant mass squared, $s_{12}$, and the meson-baryon invariant mass
squared, $s_{13}$. 
The remaining two degrees of freedom describe the orientation of the final state in the c.m. frame. The polar angle $\theta$ is the angle between the direction of the meson pair and that of the incoming photon (see Fig.~\ref{fig:kinematics}). The recoiling baryon then moves in the opposite direction. The azimuthal angle $\varphi$ describes how the plane containing the two mesons is rotated about the direction of the meson pair, measured from the baryon plane.
For real photoproduction,
the amplitudes are independent of the azimuthal orientation, $\phi$, of the electron 
scattering plane. This angle becomes relevant only in electroproduction, where
the virtual photon carries a nonzero four-momentum transfer ($k^2 \neq 0$). 

\subsection{The Hadronic Current and Dirac Amplitudes}

Following Ref.~\cite{Berends:1967vi}, we formulate the transition matrix element
for the electromagnetic production of two mesons on the nucleon as
\begin{equation}
\label{eq:transition_matrix_el}
    {\cal M}_e = \epsilon_\mu\, {J}^\mu ,
\end{equation}
where $\epsilon_\mu$ denotes the leptonic electromagnetic current,
\begin{equation}
\label{eq:def_epsilon_mu}
    \epsilon_\mu = \frac{\bar u(\bm{k}_2)\,\gamma_\mu\,u(\bm{k}_1)}{k^2},
\end{equation}
and
\begin{equation}
    \label{eq:J_mu_defined}
    {J}^\mu = \braket{p_2\,q_1\,q_2|\,{\hat J}^\mu\,|p_1}
\end{equation}
is the hadronic electromagnetic current. The definition of $\epsilon_\mu$ in
Eq.~\eqref{eq:def_epsilon_mu} is particularly convenient because it allows the
electroproduction and photoproduction processes to be treated within a unified
formalism. In the following, $\epsilon_\mu$ will be regarded as an effective
virtual-photon polarization vector. The resulting formalism is developed for
electroproduction, while the corresponding photoproduction expressions are
obtained straightforwardly by taking the real-photon limit.

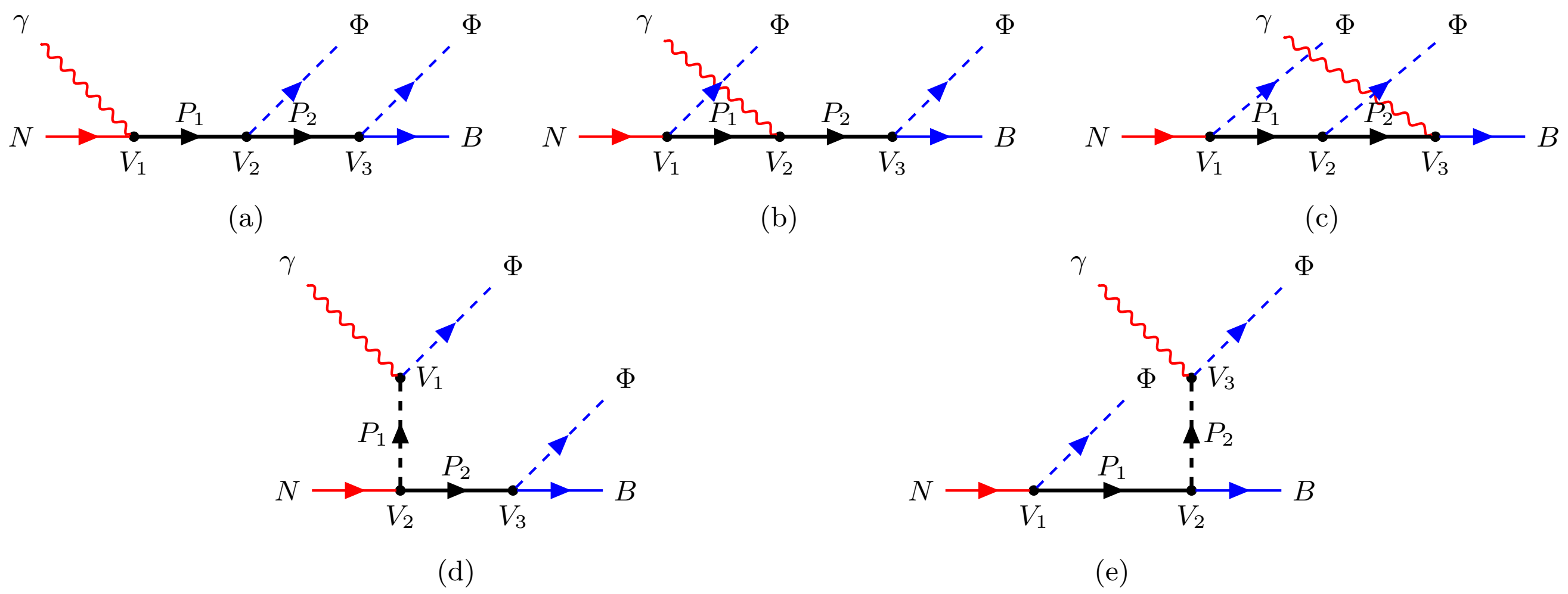
\begin{figure}
\begin{subfigure}{.33\textwidth}
    \begin{tikzpicture}[thick]
    \begin{feynman}
        \vertex at (0, 1.25) (k) {$\gamma$};
        \vertex[dot, minimum size=2.5pt, label=below:$V_1$] at (1.25, 0) (a) {};
        \vertex at (0, 0) (p1) {$N$};
        \vertex[dot, minimum size=2.5pt, label=below:$V_2$] at (2.5, 0) (b) {};
        \vertex at (3.75, 1.25) (q1) {$\Phi$};
        \vertex[dot, minimum size=2.5pt, label=below:$V_3$] at (3.75, 0) (c) {};
        \vertex at (5, 1.25) (q2) {$\Phi$};
        \vertex at (5, 0) (p2) {$B$};
    \diagram*{
        (k) -- [photon,red] (a), (p1) -- [fermion,red] (a),
        (a) -- [fermion, edge label=$P_1$, very thick] (b) -- [fermion, edge label=$P_2$, very thick] (c) -- [fermion,blue] (p2),
        (b) -- [charged scalar,blue] (q1), (c) -- [charged scalar,blue] (q2)
    };
    \end{feynman}
    \end{tikzpicture}
    \caption{} 
\end{subfigure}%
\setcounter{subfigure}{1}%
\begin{subfigure}{.33\textwidth}
    \begin{tikzpicture}[thick]
    \begin{feynman}
        \vertex  at (1, 1.25) (k) {$\gamma$};
        \vertex[dot, minimum size=2.5pt, label=below:$V_1$] at (1.25, 0) (a) {};
        \vertex at (0, 0) (p1) {$N$};
        \vertex[dot, minimum size=2.5pt, label=below:$V_2$] at (2.5, 0) (b) {};
        \vertex at (2.5, 1.25) (q1) {$\Phi$};
        \vertex[dot, minimum size=2.5pt, label=below:$V_3$] at (3.75, 0) (c) {};
        \vertex at (5, 1.25) (q2) {$\Phi$};
        \vertex at (5, 0) (p2) {$B$};
    \diagram*{
        (k) -- [photon,red] (b), (p1) -- [fermion,red] (a),
        (a) -- [fermion, edge label=$P_1$, very thick] (b) -- [fermion, edge label=$P_2$, very thick] (c) -- [fermion,blue] (p2),
        (a) -- [charged scalar,blue] (q1), (c) -- [charged scalar,blue] (q2)
    };
    \end{feynman}
    \end{tikzpicture}
    \caption{} 
\end{subfigure}
\begin{subfigure}{.33\textwidth}
    \begin{tikzpicture}[thick]
    \begin{feynman}
        \vertex at (1.85, 1.25) (k) {$\gamma$};
        \vertex[dot, minimum size=2.5pt, label=below:$V_1$] at (1.25, 0) (a) {};
        \vertex at (0, 0) (p1) {$N$};
        \vertex[dot, minimum size=2.5pt, label=below:$V_2$] at (2.5, 0) (b) {};
        \vertex at (2.75, 1.25) (q1) {$\Phi$};
        \vertex[dot, minimum size=2.5pt, label=below:$V_3$] at (3.75, 0) (c) {};
        \vertex at (4, 1.25) (q2) {$\Phi$};
        \vertex at (5, 0) (p2) {$B$};
    \diagram*{
        (k) -- [photon,red] (c), (p1) -- [fermion,red] (a),
        (a) -- [fermion, edge label=$P_1$, very thick] (b)  -- [fermion, edge label=$P_2$, very thick] (c) -- [fermion,blue] (p2),
        (a) -- [charged scalar,blue] (q1), (b) -- [charged scalar,blue] (q2)
    };
    \end{feynman}
    \end{tikzpicture}
    \caption{} 
\end{subfigure} 
\begin{subfigure}{.4\textwidth}
    \begin{tikzpicture}[thick]
    \begin{feynman}
        \vertex at (0, 2.5) (k) {$\gamma$};
        \vertex[dot, minimum size=2.5pt, label=right:$V_1$] at (1.25, 1.25) (e) {};
        \vertex[dot, minimum size=2.5pt, label=below:$V_2$] at (1.25, 0) (a) {};
        \vertex at (0, 0) (p1) {$N$};
        \vertex[dot, minimum size=2.5pt, label=below:$V_3$] at (2.5, 0) (b) {};
        \vertex at (2.5, 2.5) (q1) {$\Phi$};
        \vertex at (3.75, 1.25) (q2) {$\Phi$};
        \vertex at (3.75, 0) (p2) {$B$};
    \diagram*{
        (k) -- [photon,red] (e), (p1) -- [fermion,red] (a), (a) -- [charged scalar, edge label=$P_1$, very thick] (e),
        (a) -- [fermion, edge label=$P_2$, very thick] (b) -- [fermion,blue] (p2),
        (e) -- [charged scalar,blue] (q1), (b) -- [charged scalar,blue] (q2)
    };
    \end{feynman}
    \end{tikzpicture}
    \caption{} 
\end{subfigure} 
\begin{subfigure}{.4\textwidth}
    \begin{tikzpicture}[thick]
    \begin{feynman}
        \vertex at (1.75, 2.5) (k) {$\gamma$};
        \vertex[dot, minimum size=2.5pt, label=right:$V_3$] at (3, 1.25) (e) {};
        \vertex[dot, minimum size=2.5pt, label=below:$V_1$] at (1.25, 0) (a) {};
        \vertex at (0, 0) (p1) {$N$};
        \vertex[dot, minimum size=2.5pt, label=below:$V_2$] at (3, 0) (b) {};
        \vertex at (4.25, 2.5) (q2) {$\Phi$};
        \vertex at (2.5, 1.25) (q1) {$\Phi$};
        \vertex at (4.25, 0) (p2) {$B$};
    \diagram*{
        (k) -- [photon,red] (e), (p1) -- [fermion,red] (a), (b) -- [charged scalar, edge label'=$P_2$, very thick] (e),
        (a) -- [fermion, edge label=$P_1$, very thick] (b) -- [fermion,blue] (p2),
        (e) -- [charged scalar,blue] (q2), (a) -- [charged scalar,blue] (q1)
    };
    \end{feynman}
    \end{tikzpicture}
    \caption{} 
\end{subfigure} 
\caption{Tree-level Feynman diagrams for double-meson photo- and electroproduction $\gamma+N\to \Phi+\Phi+B$, where $\gamma$ denotes either a real or virtual photon. 
The interaction vertices and propagators are labeled by $V_i$ and $P_i$, respectively, while the thick (black) lines indicate the intermediate states. Crossed diagrams are not shown. The colors of the initial- and final-state particles are the same as those used in Fig.~\ref{fig:feynman_general}.}
\label{fig:fey_dig}
\end{figure}

Unlike the single-meson case, the electromagnetic current $J^\mu$ for double-meson production contains no explicit $\gamma_5$, even when pseudovector couplings are employed at the hadronic vertices (see Fig.~\ref{fig:fey_dig}). This follows from the fact that each tree-level production diagram contains two meson-baryon vertices, so that the two $\gamma_5$ matrices combine to yield a Dirac structure without an explicit $\gamma_5$. 


The hadronic-current operator ${\hat J}^\mu$ can be written as a linear combination of a complete set of 20 Lorentz-covariant vectors constructed from the Dirac matrices and the available four-momenta,
\begin{align} \label{eq:arus_j}
    {\hat J}^\mu = \sum\limits_{i=1}^{20}I^\mu_i \ B_i  \ ,
\end{align}
where $B_i$ are Lorentz-invariant scalar amplitudes that depend on the independent kinematic invariants, and the basis vectors $I_j^\mu$ are given by
\begin{align}
I^\mu_1 &= \gamma^\mu \ \slashed{k}\ ,         & I^\mu_6  &= \gamma^\mu\ ,                         & I^\mu_{11} &= \gamma^\mu \ \slashed{k} \slashed{q}\ , & I^\mu_{16} &= \gamma^\mu \slashed{q}\ , \nonumber \\
I^\mu_2 &= p^\mu_1\ ,                       & I^\mu_7  &= p^\mu_1 \ \slashed{k}\ ,               & I^\mu_{12} &= p^\mu_1  \ \slashed{q}\ ,               & I^\mu_{17} &= p^\mu_1 \ \slashed{k} \slashed{q}\ , \nonumber \\
I^\mu_3 &= p^\mu_2\ ,                       & I^\mu_8  &= p^\mu_2 \ \slashed{k}\ ,               & I^\mu_{13} &= p^\mu_2 \ \slashed{q}\ ,               & I^\mu_{18} &= p^\mu_2 \ \slashed{k} \slashed{q}\ ,  \\
I^\mu_4 &= q^\mu_{1}\ ,                       & I^\mu_9  &= q^\mu_{1} \ \slashed{k}\ ,               & I^\mu_{14} &= q^\mu_{1} \ \slashed{q}\ ,               & I^\mu_{19} &= q^\mu_{1} \ \slashed{k} \slashed{q}\ , \nonumber \\
I^\mu_5 &= q^\mu_{2}\ ,                       & I^\mu_{10} &= q^\mu_{2} \ \slashed{k}\ ,             & I^\mu_{15} &= q^\mu_{2} \ \slashed{q}\ ,               & I^\mu_{20} &= q^\mu_{2} \ \slashed{k} \slashed{q}\ , \nonumber
\end{align}
where we have defined 
\begin{equation}
\label{eq:q_defined}
    q = \tfrac{1}{2}(q_1-q_2).
\end{equation}

The current conservation condition on the leptonic side, $k^\mu j_\mu=0$, follows directly from the definition of the polarization vector in Eq.~\eqref{eq:def_epsilon_mu}. On the hadronic side, current conservation,
\begin{equation}
    k_\mu\, {\hat J}^\mu = 0,
\end{equation}
yields four relations among the scalar amplitudes $B_i$, 
\begin{align}
    k^2B_1  + (p_1 \cdot k) B_2  + (p_2 \cdot k) B_3 + (q_1 \cdot k) B_4  + (q_2 \cdot k) B_5 &= 0 \ , \\
    B_6  + (p_1 \cdot k) B_7  + (p_2 \cdot k) B_8 + (q_1 \cdot k) B_9  + (q_2 \cdot k) B_{10} &= 0 \ , \\
    k^2 B_{11}  + (p_1 \cdot k) B_{12}  + (p_2 \cdot k) B_{13} + (q_1 \cdot k) B_{14}  + (q_2 \cdot k) B_{15} &= 0 \ , \\
    B_{16}  + (p_1 \cdot k) B_{17}  + (p_2 \cdot k) B_{18} + (q_1 \cdot k) B_{19}  + (q_2 \cdot k) B_{20} &= 0 \ .
\end{align}
reducing the number of amplitudes by four. In the present work, we eliminate the amplitudes $B_5$, $B_6$, $B_{15}$, and $B_{16}$. Consequently, Eq.~\eqref{eq:arus_j} becomes
\begin{align} \label{eq:arus_j_2}
    {\hat J}^\mu =& \left(I_1^\mu-\frac{k^2}{q_2\cdot k}  \ I_5^\mu \right)B_1 + \left(I_2^\mu-\frac{p_1 \cdot k}{q_2 \cdot k}  \ I_5^\mu \right) B_2 + \left(I_3^\mu-\frac{p_2 \cdot k}{q_2 \cdot k} \ I_5^\mu \right) B_3 + \left( I_4^\mu - \frac{q_1 \cdot k }{q_2 \cdot k} \ I_5^\mu  \right)B_4  \nonumber \\
    &+\left( I_7^\mu - p_1 \cdot k \ I_6^\mu  \right)B_7 +\left( I_8^\mu - p_2\cdot k \ I_6^\mu  \right)B_8 +\left( I_9^\mu - q_1 \cdot k \ I_6^\mu  \right)B_9 +\left( I_{10}^\mu - q_2 \cdot k \ I_6^\mu  \right)B_{10} \nonumber \\ 
    & +\left(I_{11}^\mu-\frac{k^2}{q_2\cdot k} \ I_{15}^\mu \right)B_{11}  + \left(I_{12}^\mu-\frac{p_1 \cdot k}{q_2 \cdot k} \ I_{15}^\mu \right) B_{12} + \left(I_{13}^\mu-\frac{p_2 \cdot k}{q_2 \cdot k} \ I_{15}^\mu \right) B_{13} + \left( I_{14}^\mu - \frac{q_1 \cdot k }{q_2 \cdot k} \ I_{15}^\mu  \right)B_{14}\nonumber \\
    &  +\left( I_{17}^\mu - p_1 \cdot k \ I_{16}^\mu  \right)B_{17} +\left( I_{18}^\mu - p_2\cdot k \ I_{16}^\mu  \right)B_{18}+\left( I_{19}^\mu - q_1 \cdot k \ I_{16}^\mu  \right)B_{19} +\left( I_{20}^\mu - q_2 \cdot k \ I_{16}^\mu  \right)B_{20}\, .
\end{align}

Using Eqs.~\eqref{eq:q_defined} and \eqref{eq:arus_j_2}, together with the anticommutation relation of the Dirac matrices, $\{\gamma^\mu,\gamma^\nu\}=2g^{\mu\nu}$, we can derive
\begin{eqnarray}
\label{eq:transition_matrix_el2}
    \epsilon_\mu\, {\hat J}^\mu &=& \tfrac{1}{2} \left(\slashed{\epsilon} \slashed{k} - \slashed{k} \slashed{\epsilon}\right)  B_1 +\left( P \cdot k \ k \cdot\epsilon - P \cdot \epsilon \ k^2\right) \left( \frac{2B_1-B_2+B_3 }{ 2P \cdot k}  \right) \nonumber \\
    &&+  \left(P \cdot \epsilon \ Q \cdot k   - P \cdot k \ Q\cdot \epsilon \right) \left[\frac{k^2 \ B_1+\left(p_1- q_2\right) \cdot k \ B_2+\left(p_2+q_2\right)\cdot k~B_3+2q\cdot k~B_4}{ P \cdot k~q_2 \cdot k  } \right]  \nonumber \\
    &&+  \left(P \cdot \epsilon \ q \cdot k   - P \cdot k \ q\cdot \epsilon \right) \left[\frac{-k^2 \ B_1- p_1 \cdot k \ B_2-p_2\cdot k \ B_3-2Q \cdot k \ B_4}{ P \cdot k \ q_2 \cdot k } \right]  \nonumber \\
    &&  + \left(P \cdot \epsilon \ \slashed{k} - P \cdot k \ \slashed{\epsilon}\right)\left(B_7+B_8\right) +  \left( Q \cdot \epsilon \ \slashed{k} - Q \cdot k \ \slashed{\epsilon}  \right)\left(B_7-B_8+B_9+B_{10}\right) \nonumber \\ 
    &&+  \left(q \cdot \epsilon \ \slashed{k} - q \cdot k \ \slashed{\epsilon} \right)\left(B_9-B_{10}\right) + \frac{1}{2} \left(k \cdot \epsilon \ \slashed{k} - k^2 \ \slashed{\epsilon} \right)\left(B_8-B_7\right) \nonumber \\
    &&+ \frac{1}{2} \left(\slashed{\epsilon}  \slashed{k} \slashed{q} - \slashed{k}  \slashed{\epsilon} \slashed{q}\right) B_{11} +\left( P \cdot k \ k \cdot\epsilon \ \slashed{q} - P \cdot \epsilon \ k^2 \ \slashed{q} \right) \left( \frac{2B_{11}-B_{12}+B_{13} } {2P \cdot k} \right) \nonumber \\
    &&+  \left(P \cdot \epsilon \ Q \cdot k \ \slashed{q}   - P \cdot k \ Q\cdot \epsilon \ \slashed{q} \right) \left[\frac{k^2 \ B_{11}+\left(p_1- q_2\right) \cdot k \ B_{12}+\left(p_2+q_2\right)\cdot k \ B_{13}+2q\cdot k \ B_{14}}{ P \cdot k \ q_2 \cdot k } \right]  \nonumber \\
    &&+  \left(P \cdot \epsilon \ q \cdot k \ \slashed{q}   - P \cdot k \ q\cdot \epsilon \ \slashed{q} \right) \left[\frac{-k^2 \ B_{11}-p_1 \cdot k \ B_{12}-p_2\cdot k \ B_{13} -2Q \cdot k \ B_{14}}{ P \cdot k \ q_2 \cdot k } \right]  \nonumber \\
    & & + \left(P \cdot \epsilon \ \slashed{k} \slashed{q} - P \cdot k \ \slashed{\epsilon} \slashed{q}\right)\left(B_{17}+B_{18}\right) +  \left( Q \cdot \epsilon \ \slashed{k} \slashed{q} - Q \cdot k \ \slashed{\epsilon} \slashed{q}  \right)\left(B_{17}-B_{18}+B_{19}+B_{20}\right) \nonumber \\ 
    &&+  \left(q \cdot \epsilon \ \slashed{k} \slashed{q} - q \cdot k \ \slashed{\epsilon} \slashed{q} \right)\left(B_{19}-B_{20}\right) +  \tfrac{1}{2} \left(k \cdot \epsilon \ \slashed{k} \slashed{q} - k^2 \ \slashed{\epsilon} \slashed{q} \right)\left(B_{18}-B_{17}\right) \ ,
\end{eqnarray}
where we have introduced the four-momenta
\begin{equation}
    P= \tfrac{1}{2}(p_1+p_2) ~\qquad {\rm and}\qquad Q= \tfrac{1}{2}(q_1+q_2).
\end{equation}
By introducing 16 gauge-invariant matrices $M_i$, defined as
\begin{align}\label{eq:inv_matrices}
    M_1 &= \tfrac{1}{2}(\slashed{\epsilon} \slashed{k} - \slashed{k} \slashed{\epsilon})\ , &
    M_9 &= P \cdot k \ q \cdot \epsilon \ \slashed{q} - P \cdot \epsilon \ q \cdot k \ \slashed{q}\ , \nonumber\\
    M_2 &= P \cdot k \ Q \cdot \epsilon \ - P \cdot \epsilon \ Q \cdot k \ ,&
    M_{10} &= P \cdot \epsilon \ \slashed{k} \slashed{q} - P \cdot k \ \slashed{\epsilon} \slashed{q}\ , \nonumber\\
    M_3 &= P \cdot k \ q \cdot \epsilon \ - P \cdot \epsilon \ q \cdot k \ ,&
    M_{11} &= Q \cdot \epsilon \ \slashed{k} \slashed{q} - Q \cdot k \ \slashed{\epsilon} \slashed{q}\ , \nonumber\\
    M_4 &= P \cdot \epsilon \ \slashed{k} - P \cdot k \ \slashed{\epsilon}\ , &
    M_{12} &= q \cdot \epsilon \ \slashed{k} \slashed{q} - q \cdot k \ \slashed{\epsilon} \slashed{q}\ , \\
    M_5 &= Q \cdot \epsilon \ \slashed{k} - Q \cdot k \ \slashed{\epsilon}\ , &
    M_{13} &= \tfrac{1}{2}(k \cdot \epsilon \  \slashed{k} - k^2 \slashed{\epsilon})\ , \nonumber\\
    M_6 &= q \cdot \epsilon \ \slashed{k} - q \cdot k \ \slashed{\epsilon}\ , &
    M_{14} &= P \cdot k \ k \cdot \epsilon \  - P \cdot \epsilon \ k^2\ , \nonumber\\
    M_7 &= \tfrac{1}{2}(\slashed{\epsilon} \slashed{k} \slashed{q} - \slashed{k} \slashed{\epsilon} \slashed{q})\ , &
    M_{15} &= \tfrac{1}{2}(k \cdot \epsilon \  \slashed{k} \slashed{q} - k^2 \slashed{\epsilon} \slashed{q})\ , \nonumber\\
    M_8 &= P \cdot k \ Q \cdot \epsilon \ \slashed{q} - P \cdot \epsilon \ Q \cdot k \ \slashed{q}\ , &
    M_{16} &= P \cdot k \ k \cdot \epsilon \  \slashed{q} - P \cdot \epsilon \ k^2 \slashed{q} \ , \nonumber
\end{align}
the transition amplitude in Eq.~\eqref{eq:transition_matrix_el2} can be expressed in the more general form
\begin{align} \label{eq:inv_amp}
     \mathcal{M}= \bar{u}(p_2)\sum\limits_{i=1}^{16}  M_i \ A_i \ u(p_1) \,.
\end{align}
It is worth noting that this decomposition is the natural extension of the six gauge-invariant matrices commonly used in single-meson electroproduction. In the present case, the number of gauge-invariant matrices increases from six to sixteen because of the additional degree of freedom associated with the production of a second meson. The ten additional matrices involve the momentum $q$, which characterizes the antisymmetric combination of the two meson momenta. Note also that the last four matrices, $M_{13}$--$M_{16}$, vanish for real photons. For virtual photons they contribute to both the longitudinal and the transverse amplitudes. 
We note that these matrices are not all linearly independent. As shown in Refs.~\cite{Roberts:2004mn,Arenhovel:2014qoj}, photoproduction of two pseudoscalar mesons is described by eight independent amplitudes; the longitudinal photon polarization adds four more in electroproduction, giving twelve. Our decomposition therefore contains four redundant structures, in both photo- and electroproduction. We nevertheless retain the sixteen matrices of Eq.~\eqref{eq:inv_matrices}, because they follow directly from the covariant decomposition of the current, allow the amplitudes $A_i$ to be read off directly from the Feynman diagrams, and lead to simpler and more symmetric expressions for the observables. The redundancy does not affect the observables, since the total amplitude is unique; it only implies that the individual amplitudes $A_i$ are not uniquely determined. As in the single-meson case, the Lorentz-invariant amplitudes $A_i$ are obtained from the Feynman diagrams shown in Fig.~\ref{fig:fey_dig}. The relations between the amplitudes $A_i$ and $B_i$ are given in Appendix~\ref{app:relation_Ai_Bi}.

\subsection{Decomposition of the Dirac Amplitudes into Pauli Amplitudes}
To calculate physical observables, it is more convenient to work in Pauli space. To this end, we decompose the transition amplitude in Eq.~\eqref{eq:inv_amp} into Pauli matrices acting on two-component spinors. This decomposition is essential for exploiting the angular-momentum properties of photo- and electroproduction processes, particularly within the multipole framework \cite{Berends:1967vi}. We thus introduce the Pauli amplitudes $\mathcal{F}_j$ through
\begin{align}\label{eq:inv_amplitudes}
     \mathcal{M}
     = \bar{u}(p_2)\sum_{i=1}^{16} M_i A_i\,u(p_1)
     = \mathcal{N}\,\mathcal{X}_2^\dagger
     \sum_{j=1}^{16} L_j\,\mathcal{F}_j\,
     \mathcal{X}_1 \, ,
\end{align}
where the normalization factor is given by $\mathcal{N}=\sqrt{(E_{1}+m_1)(E_{2}+m_2)}$. The amplitudes $\mathcal{F}_j$ constitute the generalization of the six CGLN amplitudes used in single-meson electromagnetic production \cite{Chew:1957tf}. Owing to the presence of an additional meson in the final state, the number of Pauli amplitudes increases from six to sixteen, matching the number of Dirac amplitudes. 

The Pauli operators $L_j$, which are composed of the Pauli matrices and the relevant kinematic vectors, are defined as
\begin{align}\label{eq:pauli_mat}
    L_1 &= b_0, &
    L_9 &= \hat{\bm{x}}'' \cdot \bm{b}, \nonumber\\
    L_2 &= i \boldsymbol{\sigma} \cdot (\hat{\bm{k}} \times \hat{\bm{x}}'' ) b_0 \ &
    L_{10} &= i \boldsymbol{\sigma} \cdot (\hat{\bm{k}} \times \hat{\bm{x}}'' ) (\hat{\bm{x}}'' \cdot \bm{b} ), \nonumber\\
    L_3 &= i \boldsymbol{\sigma} \cdot (\hat{\bm{k}} \times \hat{\bm{Q}} ) b_0 \ &
    L_{11} &= i \boldsymbol{\sigma} \cdot (\hat{\bm{k}} \times \hat{\bm{Q}} ) (\hat{\bm{x}}'' \cdot \bm{b} ), \nonumber\\
    L_4 &= i \boldsymbol{\sigma} \cdot (\hat{\bm{x}}'' \times \hat{\bm{Q}} ) b_0, &
    L_{12} &= i \boldsymbol{\sigma} \cdot (\hat{\bm{x}}'' \times \hat{\bm{Q}} ) (\hat{\bm{x}}'' \cdot \bm{b} ), \\
    L_5 &= i \boldsymbol{\sigma} \cdot (\hat{\bm{k}} \times \bm{b} ) , &
    L_{13} &= \hat{\bm{Q}} \cdot \bm{b} , \nonumber\\
    L_6 &= i \boldsymbol{\sigma} \cdot (\hat{\bm{x}}'' \times \bm{b} ) , &
    L_{14} &= i \boldsymbol{\sigma} \cdot (\hat{\bm{k}} \times \hat{\bm{x}}'' ) (\hat{\bm{Q}} \cdot \bm{b} ), \nonumber\\
    L_7 &= i \boldsymbol{\sigma} \cdot (\hat{\bm{Q}} \times \bm{b} ) , &
    L_{15} &= i \boldsymbol{\sigma} \cdot (\hat{\bm{k}} \times \hat{\bm{Q}} ) (\hat{\bm{Q}} \cdot \bm{b} ), \nonumber\\
    L_8 &= - \boldsymbol{\sigma} \cdot (\hat{\bm{x}}'' \times \hat{\bm{Q}} ) \ \boldsymbol{\sigma} \cdot (\hat{\bm{k}} \times \bm{b} ) , &
    L_{16} &=i \boldsymbol{\sigma} \cdot (\hat{\bm{x}}'' \times \hat{\bm{Q}} ) (\hat{\bm{Q}} \cdot \bm{b} ). \nonumber
\end{align}
where, following Ref.~\cite{Berends:1967vi}, we employ $b_\mu=  \epsilon_\mu - (\boldsymbol{\epsilon} \cdot \hat{\bm{k}}/{|\bm{k}|})  \, k_\mu$, instead of $a_\mu=  \epsilon_\mu - (\epsilon_0/k_0)\, k_\mu$ introduced in Ref.~\cite{Dennery:1961zz}. This choice naturally separates the virtual-photon polarization into its transverse ($\bm{b}$) and scalar ($b_0$) components, unlike the definition of $a_\mu$, for which the scalar component vanishes identically ($a_0=0$). Furthermore, we have introduced the unit vector $\hat{\bm{x}}''$ through
\begin{align}
    \hat{\bm{q}} = \hat{\bm{x}}'' \ \sin{\vartheta} + \hat{\bm{Q}} \ \cos{\vartheta} \, .
\end{align}
By construction, $\hat{\bm{x}}''$ is orthogonal to $\hat{\bm{Q}}$, while both vectors lie in the production plane spanned by $\bm{q}_1$ and $\bm{q}_2$ (see Fig.~\ref{fig:kinematics}). Thus, as in the case of single-meson electroproduction~\cite{Berends:1967vi}, the relation between the Pauli amplitudes $\mathcal{F}_j$ and the Dirac amplitudes $A_i$ can be written in matrix form as
\begin{equation}
    \tilde{\cal F}(s,s_{12},s_{13},\theta,\varphi)=\tilde{\cal G}(s,s_{12},s_{13},\theta,\varphi)\, \tilde{A}(s,s_{12},s_{13},\theta,\varphi),
\end{equation}
where
\begin{equation}
    \tilde{\cal F}(s,s_{12},s_{13},\theta,\varphi) = \left[ \begin{array}{c}
         \mathcal{F}_1\\ \mathcal{F}_2 \\ \mathcal{F}_3 \\ \cdots \\ \cdots \\ \mathcal{F}_{16}
    \end{array} \right] \qquad {\rm and} \qquad \tilde{A}(s,s_{12},s_{13},\theta,\varphi) = \left[ \begin{array}{c}
         A_1\\ A_2 \\ A_3 \\ \cdots \\ \cdots \\ A_{16}
    \end{array} \right] ,
\end{equation}
while $\tilde{\cal G}(s,s_{12},s_{13},\theta,\varphi)$ is a $16\times 16$ matrix. The explicit relations between the Pauli and Dirac amplitudes are given in Appendix~\ref{app:Pauli_amplitudes}.


\subsection{The cross section}
\label{subsec:The_cross_section}
\begin{figure}[t]
\includegraphics[width=0.90\columnwidth]{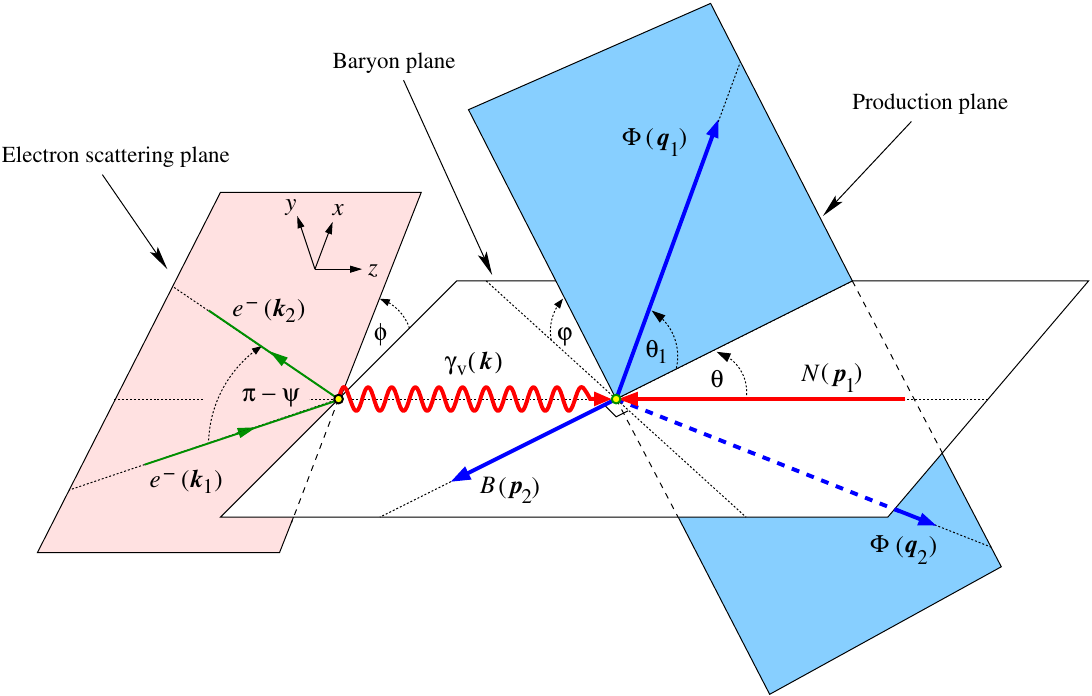} 
\caption{\label{fig:kinematics} Kinematics of double-meson virtual photoproduction on the nucleon in the $\gamma_{\rm v}$-$N$ (equivalently, the $B$-$\Phi$-$\Phi$) center-of-momentum frame. In electroproduction, the reaction is characterized by three planes: the electron-scattering, production, and baryon planes. For photoproduction, the electron-scattering plane is absent, leaving only the production and baryon planes. The straight dotted lines serve as visual guides, indicating that the corresponding momentum vectors lie within the associated planes.} 
\end{figure}

In the overall center-of-momentum frame, shown in Fig.~\ref{fig:kinematics}, the four-momenta of the incoming nucleon and photon are $p_1=(E_1,0,0,-\mathrm{p})$ and $k=(k_0,0,0,\mathrm{p})$, respectively. The outgoing baryon and mesons have four-momenta $p_2=(E_2,-\bm{p}^{\,\prime})$, $q_1=(E_{q_1},\bm{q}_1)$, and $q_2=(E_{q_2},\bm{q}_2)$, where momentum conservation requires $\bm{p}^{\,\prime}=\bm{q}_1+\bm{q}_2$.
The magnitudes of the initial- and final-state center-of-momentum momenta are
\begin{align}
    \label{eq:momenta_baryon}
    \mathrm{p}
    = \frac{\lambda^{1/2}(s,\,m_1^2,\,k^2)}{2\sqrt{s}},
    \qquad
    \mathrm{p}^{\,\prime}
    = \frac{\lambda^{1/2}(s,\,m_2^2,\,s_{12})}{2\sqrt{s}},
\end{align}
where $\lambda(x,y,z)=x^2+y^2+z^2-2xy-2xz-2yz$ is the K\"{a}ll\'{e}n
function. The relative momentum of the two mesons is defined as
$\bm{q}=(\bm{q}_1-\bm{q}_2)/2$, with individual meson energies
\begin{align}
    E_{q_1} = \frac{s_{12}+s_{13}-m_2^2-\mu_2^2}{2\sqrt{s}},
    \qquad
    E_{q_2} = \frac{s+\mu_2^2-s_{13}}{2\sqrt{s}},
\end{align}
and magnitude $\mathrm{q}=|\bm{q}|$ given by
\begin{align}
    \mathrm{q}^2
    = \frac{(2s_{13}+s_{12}-s-m_2^2-2\mu_2^2)^2
            +4s(s_{12}-2\mu_1^2-2\mu_2^2)}{16s}.
\end{align}
Furthermore, the cosine of the angle $\vartheta$ between $\bm{p}^{\,\prime}$ and $\bm{q}$ is
\begin{align}
    \cos\vartheta
    = \frac{2s(s_{13}-2\mu_1^2+\mu_2^2)
            +(s_{12}-m_2^2)^2+2(s_{13}-\mu_2^2)(s_{12}-m_2^2)-s^2}
           {8s\,\mathrm{p}^{\,\prime}\,\mathrm{q}}.
\end{align}
All kinematic quantities are thus expressed in terms of $s$, $s_{12}$, $s_{13}$,
and the particle masses, which makes them directly suitable as integration
variables for the cross section.

The physically allowed region of integration is bounded by
\begin{align}
    s_{12}^{\min} = (\mu_1+\mu_2)^2,
    \qquad
    s_{12}^{\max} = (\sqrt{s}-m_2)^2,
\end{align}
and for fixed $s_{12}$ the limits of $s_{13}$ are determined by,
\begin{align}
    s_{13}^{\max/\min}
    = \mu_1^2+m_2^2
    +\frac{(s_{12}+\mu_1^2-\mu_2^2)(s-s_{12}-m_2^2)
           \pm\lambda^{1/2}(s_{12},\mu_1^2,\mu_2^2)\,
             \lambda^{1/2}(s,m_2^2,s_{12})}
          {2s_{12}},
\end{align}
which define the boundary of the Dalitz plot.

The cross section for meson electroproduction on the nucleon with an arbitrary number of particles in the final state was derived in Ref.~\cite{Skorodumina:2015ccu}. In this subsection, we briefly summarize the derivation given in Ref.~\cite{Skorodumina:2015ccu} for the case of three final-state particles, adopting the kinematical conventions introduced in Subsec.~\ref{subsec:kinematics}. 

Using the Lorentz-invariant phase-space formalism, the differential cross section for double-meson electroproduction can be written as~\cite{Bjorken:1965sts}
\begin{equation}
\label{eq:dCS_general}
    d\sigma_e=\frac{(2\pi)^4\delta^{(4)}(k_1+p_1-k_2-q_1-q_2-p_2)}{4\sqrt{(k_1\!\cdot p_1)^2-m_e^2m_1^2}}\,\overline{|\mathcal{M}_e|^2}\,\frac{d^3k_2}{(2\pi)^3\,2E_{k_2}}\frac{d^3q_1}{(2\pi)^3\,2E_{q_1}}\frac{d^3q_2}{(2\pi)^3\,2E_{q_2}}\frac{d^3p_2}{(2\pi)^3\,2E_{p_2}},
\end{equation}
where $\overline{|\mathcal{M}_e|^2}$ denotes the squared transition amplitude averaged over the initial spins and summed over the final spins.

Within the one-photon-exchange approximation, the transition amplitude is given by Eq.~\eqref{eq:transition_matrix_el}. To adopt the normalization convention of Ref.~\cite{Skorodumina:2015ccu}, we explicitly factor out the elementary charge $e$ from both the leptonic and hadronic currents. Consequently, the coefficients $C^{(x)}$ listed in Sec.~\ref{sec:example} must be modified accordingly. In addition, the virtual-photon propagator $1/k^2$ is removed from the definition of the effective polarization vector and included explicitly in the transition amplitude. Equation~\eqref{eq:transition_matrix_el} can then be written as
\begin{eqnarray}
    \mathcal{M}_e
    &=& -\frac{4\pi\alpha}{\mathcal{Q}^2}\, j_\mu J^\mu \nonumber\\
    &=& -4\pi\alpha\, \left[\bar u(\lambda_{e_2})\gamma^\mu u(\lambda_{e_1})\right]\frac{g_{\mu\nu}}{\mathcal{Q}^2}
    J^\nu ,
\end{eqnarray}
where
\begin{equation}
    j_\mu=\bar u(\lambda_{e_2})\gamma_\mu u(\lambda_{e_1})
\end{equation}
is the leptonic electromagnetic current, with $\lambda_{e_1}$ and $\lambda_{e_2}$ denoting the helicities of the initial and final electrons, respectively. Furthermore,
\begin{equation}
    \mathcal{Q}^2=-k^2\ge0,\qquad
    k=k_1-k_2,
\end{equation}
is the squared four-momentum transfer carried by the exchanged virtual photon. 

The phase-space elements in Eq.~\eqref{eq:dCS_general} can be partly integrated and by utilizing the four-dimensional Dirac delta function the differential cross section can be written as
\begin{equation}
    \label{eq:dsigma7_1}
    d^7\sigma_e = \frac{1}{4(k_1\cdot p_1)} \frac{k_2^2\, dk_2\, d\Omega_2}{(2\pi)^3\,2E_{k_2}}\frac{(4\pi\alpha)^2}{{\cal Q}^4}\, L^{\mu\nu}\, H_{\mu\nu}\, d\Phi,
\end{equation}
with
\begin{equation}
    \label{eq:leptonic_tensor}
    L^{\mu\nu}=\tfrac{1}{2}\sum_{\lambda_{e_1}\lambda_{e_2}} \left[{\bar u}(\lambda_{e_2})\,\gamma^\mu\, u(\lambda_{e_1})\right]^*\,\left[{\bar u}(\lambda_{e_2})\,\gamma^\nu\, u(\lambda_{e_1})\right],  
\end{equation}
and
\begin{equation}
    H_{\mu\nu}=\tfrac{1}{2}\sum_{\lambda_{N}\lambda_{B}} J_{\mu}^*(\lambda_{N},\lambda_{B})\, J_\nu (\lambda_{N},\lambda_{B}) ,
\end{equation}
where we have used $J_{\mu}(\lambda_{N},\lambda_{B})\equiv\braket{\lambda_{B}|{\hat J}_{\mu}|\lambda_{N}}$, in accordance with Eq.~\eqref{eq:J_mu_defined}. Note that in Eq.~\eqref{eq:dsigma7_1} we have introduced the phase volume element $d\Phi$ which is defined as
\begin{equation}
    d\Phi= \frac{S}{32 W^2 (2\pi)^5}\, d\tau^5 ,
\end{equation}
where $S = 1/2$ for two identical mesons in the final state (e.g., $\pi^0\pi^0$) and $S = 1$ otherwise, which accounts for integrating over the full Dalitz region, and
\begin{equation}
    d\tau^5=ds_{12}\,ds_{13}\,d\cos\theta\,d\phi\,d\varphi ,
\end{equation}
which denotes the five-dimensional hadronic phase-space element obtained after eliminating the final baryon momentum using the four-momentum conservation condition.

The transversality condition,
 \begin{equation}
    k_\mu L^{\mu\nu}=k_\nu L^{\mu\nu}=0,
\end{equation}
allows the leptonic tensor in Eq.~\eqref{eq:leptonic_tensor} to be decomposed in the basis of the virtual-photon polarization vectors
$\epsilon_{\mu}(\lambda_v)$, with $\lambda_v=0,\pm1$, i.e.,
\begin{equation}
    L_{\mu\nu} = \sum_{\lambda_v,\lambda_v'}\epsilon_{\mu}(\lambda_v)\epsilon_{\nu}^{*}(\lambda_v')
    \rho_{\lambda_v\lambda_v'},
\end{equation}
where $\rho_{\lambda_v\lambda_v'}$ is the virtual-photon polarization density
matrix in the helicity basis~\cite{Schilling:1973ag}. 
Consequently, the electroproduction cross section in Eq.~\eqref{eq:dsigma7_1} can be expressed as
\begin{equation}
    \frac{d^7\sigma_e}{dE_{k_2}\,d\Omega_{e_2}\,d\tau^5} = \Gamma_v \frac{d^5\sigma_v}       {d\tau^5},
\end{equation}
with the virtual-photon flux
\begin{equation}
    \Gamma_v = \frac{\alpha}{2\pi^2} \frac{E_{k_2}}{E_{k_1}} \frac{K}{{\cal Q}^2} \frac{1}{1-\varepsilon},
\end{equation}
and the degree of transverse polarization of the virtual photon
\begin{equation}
    \varepsilon=\left[1+\frac{2\bm{k}_{\rm lab}^2}{\mathcal{Q}^2}\tan^2\!\left(\frac{\psi}{2}\right)\right]^{-1},
    \label{eq:transverse_pol_degree}
\end{equation}
where $\bm{k}_{\rm lab}$ is the three-momentum of the virtual photon and $\psi$ is the electron scattering angle, both taken in the laboratory frame. The equivalent photon energy is given by
\begin{equation}
    K = \frac{W^2-m_1^2}{2m_1},
\end{equation}
which corresponds to the laboratory energy of a real photon required to
produce a hadronic system with the same center-of-mass energy $W$. The
five-fold differential cross section for virtual-photon production then reads 
\begin{equation}
    \label{eq:five_fold_CS}
    \frac{d^5\sigma_v}{d\tau^5} = \frac{4\pi\alpha}{4Km_1} \sum_{\lambda_v,\lambda_v'}  \epsilon_{\mu}(\lambda_v) \epsilon_{\nu}^{*}(\lambda_v') \rho_{\lambda_v\lambda_v'} \frac12 \sum_{\lambda_N,\lambda_B} {J^\mu}^{*}(\lambda_N,\lambda_B) J^\nu(\lambda_N,\lambda_B) \frac12\left( \frac{1-\varepsilon}{{\cal Q}^2} \right) \frac{S}{32W^2(2\pi)^5}.
\end{equation}
Note that this expression gives the cross section for double-meson production induced by a virtual photon defined in Eq.~\eqref{eq:virtual_photo}, with the virtual-photon polarization vector $\epsilon_\mu$, and the corresponding process depicted schematically in Fig.~\ref{fig:feynman_general}. This formulation is particularly convenient for phenomenological analyses of electromagnetic double-meson production on the nucleon, in much the same way that the two-fold differential cross section $d\sigma/d\Omega_v$ is commonly used in the analysis of single-meson electro- and photoproduction on the nucleon \cite{Knochlein:1995qz}.

Both the leptonic and hadronic tensors contain 16 components. The current conservation conditions, $k_\mu j^\mu=0$ on the leptonic side (see Eq.~\eqref{eq:def_epsilon_mu}) and $k_\mu J^\mu=0$ on the hadronic side (see Eq.~\eqref{eq:J_mu_defined}), reduce the number of independent components of the leptonic and hadronic tensors from sixteen to nine. Using these relations, the scalar product of the leptonic and hadronic currents can be written as
\begin{equation}
    j_\mu J^\mu = -\left( j_xJ^x + j_yJ^y - \frac{{\cal Q}^2}{k_0^2} j_zJ^z \right) ,
\end{equation}
where the time components have been eliminated in favor of the spatial components. Note that $k_0$ denotes the virtual-photon energy in the hadronic c.m. frame, in contrast to $\bm{k}_{\rm lab}$ and $\psi$ in Eq.~\eqref{eq:transverse_pol_degree}, which are defined in the laboratory frame. Consequently, only three independent components of each current are required in the evaluation of the transition amplitude.
Following Ref.~\cite{Skorodumina:2015ccu}, the time components of the leptonic and hadronic currents are eliminated by means of the current conservation conditions. The leptonic tensor then takes the form of the $3\times3$ matrix 
\begin{equation}
    \label{eq:Lij}
    L_{ij} = \frac{\mathcal{Q}^2}{1-\varepsilon} \left( \begin{array}{ccc}
       1+\varepsilon  &  0 & -\sqrt{2\varepsilon_{\rm L}(1+\varepsilon)} \\
        0 & 1-\varepsilon  & 0 \\
        -\sqrt{2\varepsilon_{\rm L}(1+\varepsilon)} & 0 & 2\varepsilon_{\rm L} 
    \end{array} \right) ,
\end{equation}
where 
\begin{equation}
    \varepsilon_{\rm L} = \frac{{\cal Q}^2}{k_0^2}\,\varepsilon.
\end{equation}
Likewise, the hadronic tensor reduces to the $3\times3$ matrix
\begin{equation}
    \label{eq:Hij}
    H_{ij} = \frac{1}{2} \sum_{\lambda_{N}\lambda_{B}} \left(\! \begin{array}{ccc}
        J_x^*J_x & J_x^*J_y &J_x^*J_z \\
        J_y^*J_x & J_y^*J_y & J_y^*J_z \\
        J_z^*J_x & J_z^*J_y & J_z^*J_z
    \end{array} \!\right) ,
\end{equation}
where the summation runs over the spin projections of the initial nucleon and the final baryon. 

To calculate the components of the hadronic tensor in Eq.~\eqref{eq:Hij} we use Eqs.~\eqref{eq:inv_amplitudes} and \eqref{eq:pauli_mat}. For $i=x,y$ we obtain
\begin{align} \label{eq:Ji}
    {\hat J}_i=& (\boldsymbol{\sigma} \cdot \hat{\bm{k}} \ \sigma_i- \hat{k}_i ) \mathcal{F}_5 + (\boldsymbol{\sigma} \cdot \hat{\bm{x}}'' \ \sigma_i- \hat{x}''_i ) \mathcal{F}_6 + (\boldsymbol{\sigma} \cdot \hat{\bm{Q}} \ \sigma_i- \hat{Q}_i ) \mathcal{F}_7 \nonumber \\ 
    &+ i\boldsymbol{\sigma} \cdot (\hat{\bm{x}}'' \times \hat{\bm{Q}})(\boldsymbol{\sigma} \cdot \hat{\bm{k}} \ \sigma_i- \hat{k}_i ) \mathcal{F}_8 + \hat{x}''_i \ \mathcal{F}_9 + i\boldsymbol{\sigma} \cdot (\hat{\bm{k}} \times \hat{\bm{x}}'') \ \hat{x}''_i \ \mathcal{F}_{10} + i\boldsymbol{\sigma} \cdot (\hat{\bm{k}} \times \hat{\bm{Q}}) \ \hat{x}''_i \ \mathcal{F}_{11} \nonumber \\
    &+ i\boldsymbol{\sigma} \cdot (\hat{\bm{x}}'' \times \hat{\bm{Q}}) \ \hat{x}''_i \ \mathcal{F}_{12} + \hat{Q}_i \ \mathcal{F}_{13} + i\boldsymbol{\sigma} \cdot (\hat{\bm{k}} \times \hat{\bm{x}}'') \ \hat{Q}_i \ \mathcal{F}_{14} + i\boldsymbol{\sigma} \cdot (\hat{\bm{k}} \times \hat{\bm{Q}}) \ \hat{Q}_i \ \mathcal{F}_{15} \nonumber \\
    &+ i\boldsymbol{\sigma} \cdot (\hat{\bm{x}}'' \times \hat{\bm{Q}}) \ \hat{Q}_i \ \mathcal{F}_{16} \ .
\end{align}
For $i=z$, it is more convenient to use the longitudinal component $J_0$ together with the current-conservation condition,
\begin{equation}
k_0 {\hat J}_0-|\bm{k}|{\hat J}_z=0,
\end{equation}
where
\begin{equation} 
    \label{eq:J0}
    {\hat J}_0= \mathcal{F}_1 +i\boldsymbol{\sigma} \cdot (\hat{\bm{k}} \times \hat{\bm{x}}'')  \ \mathcal{F}_2 + i\boldsymbol{\sigma} \cdot (\hat{\bm{k}} \times \hat{\bm{Q}}) \ \mathcal{F}_3 + i\boldsymbol{\sigma} \cdot (\hat{\bm{x}}'' \times \hat{\bm{Q}}) \ \mathcal{F}_4  \ .
\end{equation}
The resulting components of the hadronic tensor, $J_i^*J_k$ are listed in Appendix~\ref{app:had_tensor}. 

Finally, using Eqs.~\eqref{eq:Lij} and \eqref{eq:Hij} the five-fold differential cross section in Eq.~\eqref{eq:five_fold_CS} can be reorganized in terms of the virtual-photon response functions, which are defined by the corresponding combinations of the hadronic tensor components, 
\begin{eqnarray}
    \label{eq:five_fold_CS2}
    \frac{d^5\sigma_v}{d\tau^5} &=& \frac{4\pi\alpha}{4Km_1} \sum_{\lambda_N,\lambda_B}  \frac12 \left\{ \frac{|J_x|^2+|J_y|^2}{2} + \varepsilon_{\rm L} |J_z|^2 + \varepsilon \frac{|J_x|^2-|J_y|^2}{2} - \sqrt{2\varepsilon_{\rm L}(1+\varepsilon)}~ {\rm Re} (J_x^*J_z)
    \right\} \frac{S}{32W^2(2\pi)^5} \nonumber\\
    &\equiv& \sigma_{\rm T} + \varepsilon_{\rm L}\sigma_{\rm L} + \varepsilon \left( \sigma_{\rm TT} \cos 2\phi + \sigma_{\rm TT}' \sin 2\phi \right)     + \sqrt{2\varepsilon_{\rm L}(1+\varepsilon)} \left( \sigma_{\rm TL} \cos \phi + \sigma_{\rm TL}' \sin \phi  \right)
    .
\end{eqnarray}
Note that the $\phi$ dependence is implicit in the first line, since the currents $J_{x,y,z}$ are defined in the electron-plane frame while the hadronic frame is rotated by $\phi$ about the $z$ axis; it becomes explicit in the second line after this rotation is performed (see Fig.~\ref{fig:kinematics}). For convenience, we define the common kinematical factor
\begin{equation}
\label{eq:def_F}
    F=\frac{4\pi\alpha}{4Km_1} \frac{S}{32W^2(2\pi)^5},
\end{equation}
so that the response functions in Eq.~\eqref{eq:five_fold_CS2} can be written as
\begin{subequations}
\label{eq:response_functions}
\begin{eqnarray}
    &&\sigma_{\rm T}=F\sum_{\lambda_N,\lambda_B}\frac{|J_x|^2+|J_y|^2}{4},\\
    &&\sigma_{\rm L}=F\sum_{\lambda_N,\lambda_B}\frac{|J_z|^2}{2},\\
    &&\sigma_{\rm TT} \cos 2\phi + \sigma_{\rm TT}' \sin 2\phi=F\sum_{\lambda_N,\lambda_B}\frac{|J_x|^2-|J_y|^2}{4},\\
    &&\sigma_{\rm TL} \cos \phi + \sigma_{\rm TL}' \sin \phi=-F\sum_{\lambda_N,\lambda_B}\frac{{\rm Re} (J_x^*J_z)}{2}.
\end{eqnarray}
\end{subequations}
Should they be needed, the explicit expressions for the individual response functions $\sigma_{\rm TT}$, $\sigma_{\rm TT}'$, $\sigma_{\rm TL}$, and $\sigma_{\rm TL}'$ can be derived from Eqs.~\eqref{eq:jx2-jy2} and \eqref{eq:re_jxs_jz} in Appendix~\ref{app:had_tensor}.


\section{Example: Photoproduction of Two Pions on the Proton}
\label{sec:example}
The Feynman diagrams for the general tree-level Born terms contributing to double-meson photo- and electroproduction are shown in Fig.~\ref{fig:fey_dig}. Contributions from meson and baryon resonances can be obtained by replacing the corresponding intermediate meson and baryon propagators with the appropriate resonance propagators and imposing the relevant quantum-number conservation at each vertex. For identical mesons in the final state, the corresponding crossed diagrams, which are not shown in Fig.~\ref{fig:fey_dig}, must also be included to ensure the proper symmetrization of the final state.

As an illustration of how conveniently and easily the developed formalism can be applied, in the same spirit as the one of Berends \textit{et al.} \cite{Berends:1967vi} for single-meson production, let us consider the following double-pion photoproduction processes on the proton:
\begin{align} 
    \gamma(k) + p(p_1) &\to \pi^0 (q_1) + \pi^0 (q_2) + p(p_2) \ , \\
    \gamma(k) + p(p_1) &\to \pi^+ (q_1) + \pi^0 (q_2) + n(p_2) \ ,
\end{align}
for which we have experimental data for comparison. Furthermore, this example also demonstrates an important feature of double-meson production, i.e., the individual diagrams are not gauge invariant. Gauge invariance is instead preserved by the total amplitude, as in single-meson photo- and electroproduction \cite{Haberzettl:1997jg}. 

The scalar Dirac amplitudes $A_i$ can be obtained by decomposing the scattering amplitudes into the gauge-invariant matrices defined in Eq.~\eqref{eq:inv_matrices}, using the diagrams shown in Fig.~\ref{fig:fey_dig}. For both processes, we consider only the Born-term contributions; that is, the amplitudes $A_i$ are calculated from the non-resonant diagrams. Since the purpose of this calculation is primarily to illustrate the application of the formalism developed in this work, we consider the Born-term treatment sufficient for the present study. A more complete treatment including meson and baryon resonance contributions is beyond the scope of the present study and will be considered in future investigations.
The invariant amplitudes are constructed using pseudoscalar coupling, and therefore contact-term contributions are absent. For simplicity, all intermediate nucleon propagators, whether proton or neutron,
are assigned a common mass equal to the isospin average of the doublet,
\begin{align}
    m \equiv \frac{m_p + m_n}{2} = 938.92~\mathrm{MeV}.
\end{align}
The external nucleon masses are kept at their physical values. We have verified numerically that this approximation changes our results by only about 1--2\%, which we consider acceptable for the present exploratory study. Since the amplitude is evaluated through the gauge-invariant matrices $M_i$, current conservation remains exact.

\subsection{$\gamma p \to \pi^0 \pi^0 p$}
\label{sec:res_pi0_pi0}

\begin{figure}
\begin{subfigure}{.33\textwidth}
    \begin{tikzpicture}[thick]
    \begin{feynman}
        \vertex at (0, 1.25) (k) {$\gamma$};
        \vertex[dot, minimum size=2.5pt] at (1.25, 0) (a) {};
        \vertex at (0, 0) (p1) {$p$};
        \vertex[dot, minimum size=2.5pt] at (2.5, 0) (b) {};
        \vertex at (3.75, 1.25) (q1) {$\pi^0$};
        \vertex[dot, minimum size=2.5pt] at (3.75, 0) (c) {};
        \vertex at (5, 1.25) (q2) {$\pi^0$};
        \vertex at (5, 0) (p2) {$p$};
    \diagram*{
        (k) -- [photon, red] (a), (p1) -- [fermion, red, very thick] (a),
        (a) -- [fermion, edge label'=$p$, very thick] (b) -- [fermion, edge label'=$p$ , very thick] (c) -- [fermion,blue, very thick] (p2),
        (b) -- [charged scalar,blue,  thick] (q1), (c) -- [charged scalar,blue, ,  thick] (q2)
    };
    \end{feynman}
    \end{tikzpicture}
    \caption{}
    \label{fig:fey_dig_1a}
\end{subfigure}%
\setcounter{subfigure}{1}%
\begin{subfigure}{.33\textwidth}
    \begin{tikzpicture}[thick]
    \begin{feynman}
        \vertex  at (1, 1.25) (k) {$\gamma$};
        \vertex[dot, minimum size=2.5pt] at (1.25, 0) (a) {};
        \vertex at (0, 0) (p1) {$p$};
        \vertex[dot, minimum size=2.5pt] at (2.5, 0) (b) {};
        \vertex at (2.5, 1.25) (q1) {$\pi^0$};
        \vertex[dot, minimum size=2.5pt] at (3.75, 0) (c) {};
        \vertex at (5, 1.25) (q2) {$\pi^0$};
        \vertex at (5, 0) (p2) {$p$};
    \diagram*{
        (k) -- [photon,red] (b), (p1) -- [fermion,red, very thick] (a),
        (a) -- [fermion, edge label'=$p$, very thick] (b) -- [fermion, edge label'=$p$, very thick] (c) -- [fermion,blue, very thick] (p2),
        (a) -- [charged scalar,blue, thick] (q1), (c) -- [charged scalar,blue, thick] (q2)
    };
    \end{feynman}
    \end{tikzpicture}
    \caption{}
    \label{fig:fey_dig_1b}
\end{subfigure}
\begin{subfigure}{.33\textwidth}
    \begin{tikzpicture}[thick]
    \begin{feynman}
        \vertex at (1.5, 1.25) (k) {$\gamma$};
        \vertex[dot, minimum size=2.5pt] at (1.25, 0) (a) {};
        \vertex at (0, 0) (p1) {$p$};
        \vertex[dot, minimum size=2.5pt] at (2.5, 0) (b) {};
        \vertex at (2.5, 1.25) (q1) {$\pi^0$};
        \vertex[dot, minimum size=2.5pt] at (3.75, 0) (c) {};
        \vertex at (3.75, 1.25) (q2) {$\pi^0$} ;
        \vertex at (5, 0) (p2) {$p$};
    \diagram*{
        (k) -- [photon,red] (c), (p1) -- [fermion,red, very thick] (a),
        (a) -- [fermion, edge label'=$p$, very thick] (b)  -- [fermion, edge label'=$p$, very thick] (c) -- [fermion,blue, very thick] (p2),
        (a) -- [charged scalar,blue, thick] (q1), (b) -- [charged scalar,blue, thick] (q2)
    };
    \end{feynman}
    \end{tikzpicture}
    \caption{}
    \label{fig:fey_dig_1c}
\end{subfigure}
\caption{Same as in Fig.~\ref{fig:fey_dig}, but for the process of $\gamma p \to \pi^0\pi^0p$. The diagram in (\subref{fig:fey_dig_1a}) represents the $s$-channel, while (\subref{fig:fey_dig_1b}) and (\subref{fig:fey_dig_1c}) are the $u$-channel Born terms. The corresponding crossed diagrams are not shown.}
\label{fig:fey_dig_1}
\end{figure}
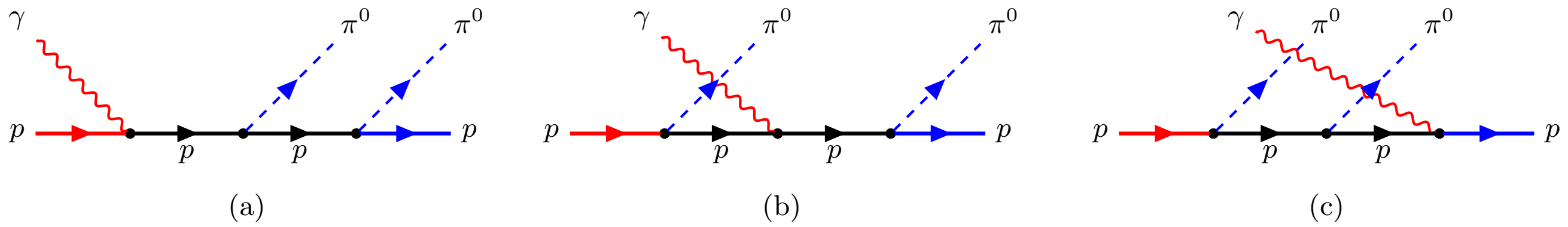

The corresponding Born diagrams for $\gamma p \to \pi^0 \pi^0 p$ process are shown in Fig. \ref{fig:fey_dig_1}. The invariant amplitudes obtained from the diagrams shown in Fig. \ref{fig:fey_dig_1}(\subref{fig:fey_dig_1a}), (\subref{fig:fey_dig_1b}), and (\subref{fig:fey_dig_1c}), respectively, are given by
\begin{align}
    \mathcal{M}^{(s_{1})} =~& \frac{-ie \ g_{\pi^0 p p}  \ g_{\pi^0 p p} }{[(p_2+q_2)^2-m^2][(p_1+k)^2-m^2]} \ \bar{u} (p_2)  \ (\slashed{p}_2 + \slashed{q}_2 - m) \nonumber \\ 
    & \times  (\slashed{p}_1 + \slashed{k} + m)  \left[ \slashed{\epsilon}  - \frac{\kappa_p}{4m}  (\slashed{\epsilon} \slashed{k} -\slashed{k} \slashed{\epsilon} )  \right] u (p_1)  \ ,  \label{eq:invamp_1_s}\\
    \mathcal{M}^{(u_{11})} =~& \frac{-ie \ g_{\pi^0 p p}   \ g_{\pi^0 p p}   }{[(p_2+q_2)^2-m^2][(p_1-q_1)^2-m^2]} \ \bar{u} (p_2)  \ (\slashed{p}_2 + \slashed{q}_2 - m) \nonumber \\
    &\times \left[ \slashed{\epsilon} + \frac{\kappa_p }{4m}  (\slashed{\epsilon} \slashed{k} - \slashed{k} \slashed{\epsilon} )  \right] (\slashed{p}_1 -\slashed{q}_1 - m) \ u (p_1) \ , \label{eq:invamp_1_u1} \\
    \mathcal{M}^{(u_{12})}=~& \frac{-ie \ g_{\pi^0 p p} \ g_{\pi^0 p p}}{[(p_2-k)^2-m^2][(p_1-q_1)^2-m^2]} \ \bar{u} (p_2)  \left[ \slashed{\epsilon}  - \frac{\kappa_p}{4m}  (\slashed{\epsilon} \slashed{k} - \slashed{k} \slashed{\epsilon})  \right] \nonumber \\
    & \times (\slashed{p}_2 - \slashed{k} + m)   (\slashed{p}_1 -\slashed{q}_1 -m) \ u (p_1)  \label{eq:invamp_1_u2}\ ,
\end{align}
where $\kappa_p$ denotes the anomalous magnetic moment of the proton. It is clear that the individual diagrams are not gauge invariant by themselves because each contains a non-gauge-invariant electric term arising from the electromagnetic nucleon vertex. This can be verified directly by replacing $\slashed{\epsilon}$ with $\slashed{k}$ in each amplitude. However, gauge invariance is restored when all three diagrams are combined to form the total amplitude. Explicitly, this can be demonstrated as follows: \begin{align}
    \MoveEqLeft 
    \left[ \mathcal{M}^{(s_1)}_{\rm el} + \mathcal{M}^{(u_{11})}_{\rm el} + \mathcal{M}^{(u_{12})}_{\rm el} \right]_{\epsilon^\mu\to k^\mu} \nonumber\\
    =& -ie \ g_{\pi^0 p p} \ g_{\pi^0 p p} \ \bar{u} (p_2)\biggl\{ \frac{(\slashed{p}_2 + \slashed{q}_2 - m)(\slashed{p}_1 + \slashed{k} + m) \slashed{k}}{[(p_2+q_2)^2-m^2][(p_1+k)^2-m^2]} \nonumber \\
    &+ \frac{(\slashed{p}_2 + \slashed{q}_2 - m)\slashed{k}(\slashed{p}_1 - \slashed{q}_1 - m) }{[(p_2+q_2)^2-m^2][(p_1-q_1)^2-m^2]} \nonumber \\
    &+ \frac{\slashed{k}(\slashed{p}_2 - \slashed{k} + m)(\slashed{p}_1 - \slashed{q}_1 - m) }{[(p_2-k)^2-m^2][(p_1-q_1)^2-m^2]} \biggr\} u (p_1) \nonumber \\
    =& -ie \ g_{\pi^0 p p} \ g_{\pi^0 p p} \ \bar{u} (p_2) \biggl\{ \frac{(\slashed{p}_2 + \slashed{q}_2 - m)}{(p_2+q_2)^2-m^2} \nonumber \\
    & \times \biggl[ \frac{(\slashed{p}_1 + \slashed{k} + m) ( \slashed{p}_1 + \slashed{k} - m)}{(p_1+k)^2-m^2} - \frac{(\slashed{p}_1 + \slashed{k} + m) ( \slashed{p}_1  - m)}{(p_1+k)^2-m^2} \nonumber \\
    &- \frac{(\slashed{p}_2 + \slashed{q}_2 - \slashed{k} + m)(\slashed{p}_1 - \slashed{q}_1 - m) }{(p_1-q_1)^2-m^2} + \frac{(\slashed{p}_2 + \slashed{q}_2 + m)(\slashed{p}_1 - \slashed{q}_1 - m) }{(p_1-q_1)^2-m^2} \biggr] \nonumber \\
    &+ \frac{\slashed{k}(\slashed{p}_2 - \slashed{k} + m)(\slashed{p}_1 - \slashed{q}_1 - m) }{[(p_2-k)^2-m^2][(p_1-q_1)^2-m^2]} \biggr\} u (p_1) \nonumber \\
    =& ~0 \ . \label{eq:gi_proof_1_1}
\end{align}
We can also construct the invariant amplitudes from the crossed diagrams which are not shown in Fig. \ref{fig:fey_dig_1}, in which we switch mesons $q_1$ and $q_2$ from the invariant amplitudes Eqs. \eqref{eq:invamp_1_s}, \eqref{eq:invamp_1_u1}, and \eqref{eq:invamp_1_u2}, so we obtain
\begin{align}
    \mathcal{M}^{(s_{2})} =~& \frac{-ie \ g_{\pi^0 p p}  \ g_{\pi^0 p p} }{[(p_2+q_1)^2-m^2][(p_1+k)^2-m^2]} \ \bar{u} (p_2)  \ (\slashed{p}_2 + \slashed{q}_1 - m) \nonumber \\ 
    & \times  (\slashed{p}_1 + \slashed{k} + m)  \left[ \slashed{\epsilon}  - \frac{\kappa_p}{4m}  (\slashed{\epsilon} \slashed{k} -\slashed{k} \slashed{\epsilon} )  \right] u (p_1) \label{eq:invamp_1_s2} \ ,\\
    \mathcal{M}^{(u_{21})} =~& \frac{-ie \ g_{\pi^0 p p}   \ g_{\pi^0 p p}   }{[(p_2+q_1)^2-m^2][(p_1-q_2)^2-m^2]} \ \bar{u} (p_2)  \ (\slashed{p}_2 + \slashed{q}_1 - m) \nonumber \\
    &\times \left[ \slashed{\epsilon} + \frac{\kappa_p }{4m}  (\slashed{\epsilon} \slashed{k} - \slashed{k} \slashed{\epsilon} )  \right] (\slashed{p}_1 -\slashed{q}_2 - m) \ u (p_1) \label{eq:invamp_1_u21} \ ,\\
    \mathcal{M}^{(u_{22})}=~& \frac{-ie \ g_{\pi^0 p p} \ g_{\pi^0 p p}}{[(p_2-k)^2-m^2][(p_1-q_2)^2-m^2]} \ \bar{u} (p_2)  \left[ \slashed{\epsilon}  - \frac{\kappa_p}{4m}  (\slashed{\epsilon} \slashed{k} - \slashed{k} \slashed{\epsilon})  \right] \nonumber \\
    & \times (\slashed{p}_2 - \slashed{k} + m)   (\slashed{p}_1 -\slashed{q}_2 -m) \ u (p_1)  \label{eq:invamp_1_u22} \ .
\end{align}
Note that the individual diagrams for this case are also not gauge invariant by themselves. Identical to Eq.~\eqref{eq:gi_proof_1_1}, gauge-invariance is restored by combining all of the crossed diagrams to form the total amplitude, demonstrated as follows:
\begin{align}
    \MoveEqLeft 
    \left[ \mathcal{M}^{(s_2)}_{\rm el} + \mathcal{M}^{(u_{21})}_{\rm el} + \mathcal{M}^{(u_{22})}_{\rm el} \right]_{\epsilon^\mu\to k^\mu} \nonumber\\
    =& -ie \ g_{\pi^0 p p} \ g_{\pi^0 p p} \ \bar{u} (p_2)\biggl\{ \frac{(\slashed{p}_2 + \slashed{q}_1 - m)(\slashed{p}_1 + \slashed{k} + m) \slashed{k}}{[(p_2+q_1)^2-m^2][(p_1+k)^2-m^2]} \nonumber \\
    &+ \frac{(\slashed{p}_2 + \slashed{q}_1 - m)\slashed{k}(\slashed{p}_1 - \slashed{q}_2 - m) }{[(p_2+q_1)^2-m^2][(p_1-q_2)^2-m^2]} \nonumber \\
    &+ \frac{\slashed{k}(\slashed{p}_2 - \slashed{k} + m)(\slashed{p}_1 - \slashed{q}_2 - m) }{[(p_2-k)^2-m^2][(p_1-q_2)^2-m^2]} \biggr\} u (p_1) \nonumber \\
    =& -ie \ g_{\pi^0 p p} \ g_{\pi^0 p p} \ \bar{u} (p_2) \biggl\{ \frac{(\slashed{p}_2 + \slashed{q}_1 - m)}{(p_2+q_1)^2-m^2} \nonumber \\
    & \times \biggl[ \frac{(\slashed{p}_1 + \slashed{k} + m) ( \slashed{p}_1 + \slashed{k} - m)}{(p_1+k)^2-m^2} - \frac{(\slashed{p}_1 + \slashed{k} + m) ( \slashed{p}_1  - m)}{(p_1+k)^2-m^2} \nonumber \\
    &- \frac{(\slashed{p}_2 + \slashed{q}_1 - \slashed{k} + m)(\slashed{p}_1 - \slashed{q}_2 - m) }{(p_1-q_2)^2-m^2} + \frac{(\slashed{p}_2 + \slashed{q}_1 + m)(\slashed{p}_1 - \slashed{q}_2 - m) }{(p_1-q_2)^2-m^2} \biggr] \nonumber \\
    &+ \frac{\slashed{k}(\slashed{p}_2 - \slashed{k} + m)(\slashed{p}_1 - \slashed{q}_2 - m) }{[(p_2-k)^2-m^2][(p_1-q_2)^2-m^2]} \biggr\} u (p_1) \nonumber \\
    =& ~0 \ . \label{eq:gi_proof_1_2}
\end{align}
The total Dirac amplitudes $A_i$ obtained from the decomposition of Eqs. \eqref{eq:invamp_1_s}, \eqref{eq:invamp_1_u1},  \eqref{eq:invamp_1_u2}, \eqref{eq:invamp_1_s2}, \eqref{eq:invamp_1_u21}, and \eqref{eq:invamp_1_u22}  are
\begin{align}
    A^{(\mathbf{A})}_1 =& -\frac{\kappa_p}{2m}\Biggl[\left(\frac{k^2+2p_1\cdot k}{2} \right)\big(C^{(s_1)}+C^{(s_2)}\big) -\big(q_2\cdot k+2q \cdot Q - q_1 \cdot q_2 - 2p_1 \cdot q\big)C^{(u_{11})} \nonumber\\
    &-\big(q_1\cdot k-2q \cdot Q - q_1 \cdot q_2 +2p_1 \cdot q\big)C^{(u_{21})} - \left(\frac{2p_2\cdot k-k^2}{2}\right)\big(C^{(u_{12})}+C^{(u_{22})}\big) \Biggr] \ , \\
    A^{(\mathbf{A})}_2 =& \frac{\kappa_p}{m}\left[\big(C^{(u_{11})}+C^{(u_{21})}\big) +\frac{q\cdot k}{P\cdot k}\Big(2C^{(s_1)}-2C^{(s_2)}-C^{(u_{11})}+C^{(u_{21})}\Big)\right] \ , \\
    A^{(\mathbf{A})}_3 =& -\frac{\kappa_p}{m\,P\cdot k}\Big[2(P\cdot k+Q\cdot k)\big(C^{(s_1)}-C^{(s_2)}\big) -(Q\cdot k-P\cdot k)\big(C^{(u_{11})}-C^{(u_{21})}\big)\Big] \ , \\
    A^{(\mathbf{A})}_5 =& -(1+\kappa_p)\big(C^{(u_{11})}+C^{(u_{21})}\big) \ , \\
    A^{(\mathbf{A})}_6 =& 2(1+\kappa_p)\big(C^{(s_1)}-C^{(s_2)}\big) +(1+\kappa_p)\big(C^{(u_{11})}-C^{(u_{21})}\big) \ , \\
    A^{(\mathbf{A})}_7 =& (1+\kappa_p)\big(C^{(s_1)}-C^{(s_2)}+C^{(u_{11})}+C^{(u_{12})}-C^{(u_{21})}-C^{(u_{22})}\big) \ , \\
    A^{(\mathbf{A})}_8 =& \frac{2(P\cdot k+q\cdot k)\big(C^{(s_1)}-C^{(s_2)}-C^{(u_{21})}\big) +2(P\cdot k-q\cdot k)\big(C^{(u_{11})}+C^{(u_{12})}-C^{(u_{22})}\big)}{P\cdot k\;q_2\cdot k} \ , \\
    A^{(\mathbf{A})}_9 =& \frac{-2(P\cdot k+Q\cdot k)\big(C^{(s_1)}-C^{(s_2)}-C^{(u_{21})}\big) -2(P\cdot k-Q\cdot k)\big(C^{(u_{11})}+C^{(u_{12})}-C^{(u_{22})}\big)}{P\cdot k\;q_2\cdot k} \ , \\
    A^{(\mathbf{A})}_{10} =& -\frac{\kappa_p}{m}\big(C^{(s_1)}-C^{(s_2)}+C^{(u_{11})}+C^{(u_{12})}-C^{(u_{21})}-C^{(u_{22})}\big) \ , \\
    A^{(\mathbf{A})}_{11} =& \frac{\kappa_p}{m}\big(C^{(u_{12})}-C^{(u_{22})}-C^{(s_1)}+C^{(s_2)}\big) \ , \\
    A^{(\mathbf{A})}_{12} =& \frac{\kappa_p}{m}\big(C^{(u_{11})}+C^{(u_{21})}\big) \ , \\
    A^{(\mathbf{A})}_4 =& A^{(\mathbf{A})}_{13} =A^{(\mathbf{A})}_{14} = A^{(\mathbf{A})}_{15} = A^{(\mathbf{A})}_{16} = 0 \ ,
\end{align}
where the factors $C^{(s_i)}$ and $C^{(u_{ij})}$, with $i, j \in \{1, 2\}$, represent the product of coupling constants and propagators from the corresponding diagrams, i.e., 
\begin{align}
    C^{(s_1)}=~& \frac{-ie \ g_{\pi^0 p p} \ g_{\pi^0 p p}}{[(p_2+q_2)^2-m^2][(p_1+k)^2-m^2]} \ , \\
    C^{(s_2)}=~& \frac{-ie \ g_{\pi^0 p p} \ g_{\pi^0 p p}}{[(p_2+q_1)^2-m^2][(p_1+k)^2-m^2]} \ , \\
    C^{(u_{11})} =~& \frac{-ie \ g_{\pi^0 p p} \ g_{\pi^0 p p}}{[(p_2+q_2)^2-m^2][(p_1-q_1)^2-m^2]} \ ,  \\
    C^{(u_{12})} =~& \frac{-ie \ g_{\pi^0 p p} \ g_{\pi^0 p p}}{[(p_2-k)^2-m^2][(p_1-q_1)^2-m^2]} \ , \\
    C^{(u_{21})} =~& \frac{-ie \ g_{\pi^0 p p} \ g_{\pi^0 p p}}{[(p_2+q_1)^2-m^2][(p_1-q_2)^2-m^2]} \ ,  \\
    C^{(u_{22})} =~& \frac{-ie \ g_{\pi^0 p p} \ g_{\pi^0 p p}}{[(p_2-k)^2-m^2][(p_1-q_2)^2-m^2]} \ .
\end{align}

\subsection{$\gamma p \to \pi^+ \pi^0 n$}
\label{sec:res_pip_pi0}

\begin{figure}
\begin{subfigure}{.33\textwidth}
    \begin{tikzpicture}[thick]
    \begin{feynman}
        \vertex at (0, 1.25) (k) {$\gamma$};
        \vertex[dot, minimum size=2.5pt] at (1.25, 0) (a) {};
        \vertex at (0, 0) (p1) {$p$};
        \vertex[dot, minimum size=2.5pt] at (2.5, 0) (b) {};
        \vertex at (3.75, 1.25) (q1) {$\pi^+$};
        \vertex[dot, minimum size=2.5pt] at (3.75, 0) (c) {};
        \vertex at (5, 1.25) (q2) {$\pi^0$};
        \vertex at (5, 0) (p2) {$n$};
    \diagram*{
        (k) -- [photon,red] (a), (p1) -- [fermion,red] (a),
        (a) -- [fermion, edge label'=$p$, very thick] (b) -- [fermion, edge label'=$n$, very thick] (c) -- [fermion,blue] (p2),
        (b) -- [charged scalar,blue] (q1), (c) -- [charged scalar,blue] (q2)
    };
    \end{feynman}
    \end{tikzpicture}
    \caption{}
    \label{fig:fey_dig_2a}
\end{subfigure}%
\setcounter{subfigure}{1}%
\begin{subfigure}{.3\textwidth}
    \begin{tikzpicture}[thick]
    \begin{feynman}
        \vertex  at (1, 1.25) (k) {$\gamma$};
        \vertex[dot, minimum size=2.5pt] at (1.25, 0) (a) {};
        \vertex at (0, 0) (p1) {$p$};
        \vertex[dot, minimum size=2.5pt] at (2.5, 0) (b) {};
        \vertex at (2.5, 1.25) (q1) {$\pi^+$};
        \vertex[dot, minimum size=2.5pt] at (3.75, 0) (c) {};
        \vertex at (5, 1.25) (q2) {$\pi^0$};
        \vertex at (5, 0) (p2) {$n$};
    \diagram*{
        (k) -- [photon,red] (b), (p1) -- [fermion,red] (a),
        (a) -- [fermion, edge label'=$n$, very thick] (b) -- [fermion, edge label'=$n$, very thick] (c) -- [fermion,blue] (p2),
        (a) -- [charged scalar,blue] (q1), (c) -- [charged scalar,blue] (q2)
    };
    \end{feynman}
    \end{tikzpicture}
    \caption{}
    \label{fig:fey_dig_2b}
\end{subfigure}
\begin{subfigure}{.3\textwidth}
    \begin{tikzpicture}[thick]
    \begin{feynman}
        \vertex at (1.5, 1.25) (k) {$\gamma$};
        \vertex[dot, minimum size=2.5pt] at (1.25, 0) (a) {};
        \vertex at (0, 0) (p1) {$p$};
        \vertex[dot, minimum size=2.5pt] at (2.5, 0) (b) {};
        \vertex at (2.5, 1.25) (q1) {$\pi^+$};
        \vertex[dot, minimum size=2.5pt] at (3.75, 0) (c) {};
        \vertex at (3.75, 1.25) (q2) {$\pi^0$};
        \vertex at (5, 0) (p2) {$n$};
    \diagram*{
        (k) -- [photon,red] (c), (p1) -- [fermion,red] (a),
        (a) -- [fermion, edge label'=$n$, very thick] (b)  -- [fermion, edge label'=$n$, very thick] (c) -- [fermion,blue] (p2),
        (a) -- [charged scalar,blue] (q1), (b) -- [charged scalar,blue] (q2)
    };
    \end{feynman}
    \end{tikzpicture}
    \caption{}
    \label{fig:fey_dig_2c}
\end{subfigure} 
\begin{subfigure}{.3\textwidth}
    \begin{tikzpicture}[thick]
    \begin{feynman}
        \vertex at (0, 2.5) (k) {$\gamma$};
        \vertex[dot, minimum size=2.5pt] at (1.25, 1.25) (e) {};
        \vertex[dot, minimum size=2.5pt] at (1.25, 0) (a) {};
        \vertex at (0, 0) (p1) {$p$};
        \vertex[dot, minimum size=2.5pt] at (2.5, 0) (b) {};
        \vertex at (2.5, 2.5) (q1) {$\pi^+$};
        \vertex at (3.75, 1.25) (q2) {$\pi^0$};
        \vertex at (3.75, 0) (p2) {$n$};
    \diagram*{
        (k) -- [photon,red] (e), (p1) -- [fermion,red] (a), (a) -- [charged scalar, edge label'=$\pi^+$, very thick] (e),
        (a) -- [fermion, edge label'=$n$, very thick] (b) -- [fermion,blue] (p2),
        (e) -- [charged scalar,blue] (q1), (b) -- [charged scalar,blue] (q2)
    };
    \end{feynman}
    \end{tikzpicture}
    \caption{}
    \label{fig:fey_dig_2d}
\end{subfigure}
\begin{subfigure}{.3\textwidth}
    \begin{tikzpicture}[thick]
    \begin{feynman}
        \vertex at (0, 1.25) (k) {$\gamma$};
        \vertex[dot, minimum size=2.5pt] at (1.25, 0) (a) {};
        \vertex at (0, 0) (p1) {$p$};
        \vertex[dot, minimum size=2.5pt] at (2.5, 0) (b) {};
        \vertex at (3.75, 1.25) (q1) {$\pi^0$};
        \vertex[dot, minimum size=2.5pt] at (3.75, 0) (c) {};
        \vertex at (5, 1.25) (q2) {$\pi^+$};
        \vertex at (5, 0) (p2) {$n$};
    \diagram*{
        (k) -- [photon,red] (a), (p1) -- [fermion,red] (a),
        (a) -- [fermion, edge label'=$p$, very thick] (b) -- [fermion, edge label'=$p$, very thick] (c) -- [fermion,blue] (p2),
        (b) -- [charged scalar,blue] (q1), (c) -- [charged scalar,blue] (q2)
    };
    \end{feynman}
    \end{tikzpicture}
    \caption{}
    \label{fig:fey_dig_2e}
\end{subfigure}
\begin{subfigure}{.3\textwidth}
    \begin{tikzpicture}[thick]
    \begin{feynman}
        \vertex  at (1, 1.25) (k) {$\gamma$};
        \vertex[dot, minimum size=2.5pt] at (1.25, 0) (a) {};
        \vertex at (0, 0) (p1) {$p$};
        \vertex[dot, minimum size=2.5pt] at (2.5, 0) (b) {};
        \vertex at (2.5, 1.25) (q1) {$\pi^0$};
        \vertex[dot, minimum size=2.5pt] at (3.75, 0) (c) {};
        \vertex at (5, 1.25) (q2) {$\pi^+$};
        \vertex at (5, 0) (p2) {$n$};
    \diagram*{
        (k) -- [photon,red] (b), (p1) -- [fermion,red] (a),
        (a) -- [fermion, edge label'=$p$, very thick] (b) -- [fermion, edge label'=$p$, very thick] (c) -- [fermion,blue] (p2),
        (a) -- [charged scalar,blue] (q1), (c) -- [charged scalar,blue] (q2)
    };
    \end{feynman}
    \end{tikzpicture}
    \caption{}
    \label{fig:fey_dig_2f}
\end{subfigure}
\begin{subfigure}{.4\textwidth}
    \begin{tikzpicture}[thick]
    \begin{feynman}
        \vertex at (1.5, 1.25) (k) {$\gamma$};
        \vertex[dot, minimum size=2.5pt] at (1.25, 0) (a) {};
        \vertex at (0, 0) (p1) {$p$};
        \vertex[dot, minimum size=2.5pt] at (2.5, 0) (b) {};
        \vertex at (2.5, 1.25) (q1) {$\pi^0$};
        \vertex[dot, minimum size=2.5pt] at (3.75, 0) (c) {};
        \vertex at (3.75, 1.25) (q2) {$\pi^+$};
        \vertex at (5, 0) (p2) {$n$};
    \diagram*{
        (k) -- [photon,red] (c), (p1) -- [fermion,red] (a),
        (a) -- [fermion, edge label'=$p$, very thick] (b)  -- [fermion, edge label'=$n$, very thick] (c) -- [fermion,blue] (p2),
        (a) -- [charged scalar,blue] (q1), (b) -- [charged scalar,blue] (q2)
    };
    \end{feynman}
    \end{tikzpicture}
    \caption{}
    \label{fig:fey_dig_2g}
\end{subfigure}
\begin{subfigure}{.4\textwidth}
    \begin{tikzpicture}[thick]
    \begin{feynman}
        \vertex at (1.75, 2.5) (k) {$\gamma$};
        \vertex[dot, minimum size=2.5pt] at (3, 1.25) (e) {};
        \vertex[dot, minimum size=2.5pt] at (1.25, 0) (a) {};
        \vertex at (0, 0) (p1) {$p$};
        \vertex[dot, minimum size=2.5pt] at (3, 0) (b) {};
        \vertex at (4.25, 2.5) (q2) {$\pi^+$};
        \vertex at (2.5, 1.25) (q1) {$\pi^0$};
        \vertex at (4.25, 0) (p2) {$n$};
    \diagram*{
        (k) -- [photon,red] (e), (p1) -- [fermion,red] (a), (b) -- [charged scalar, edge label'=$\pi^+$, very thick] (e),
        (a) -- [fermion, edge label'=$p$, very thick] (b) -- [fermion,blue] (p2),
        (e) -- [charged scalar,blue] (q2), (a) -- [charged scalar,blue] (q1)
    };
    \end{feynman}
    \end{tikzpicture}
    \caption{}
    \label{fig:fey_dig_2h}
\end{subfigure}
\caption{Tree-level Feynman diagrams for the process $\gamma p \to \pi^+ \pi^0 n$. Panels (\subref{fig:fey_dig_2a}) and (\subref{fig:fey_dig_2e}) represent the $s$-channel Born terms; panels (\subref{fig:fey_dig_2b}), (\subref{fig:fey_dig_2c}), (\subref{fig:fey_dig_2f}), and (\subref{fig:fey_dig_2g}) represent the $u$-channel Born terms; and panels (\subref{fig:fey_dig_2d}) and (\subref{fig:fey_dig_2h}) represent the $t$-channel Born terms. Two attachment configurations are considered: in panels (\subref{fig:fey_dig_2a})--(\subref{fig:fey_dig_2d}), the $\pi^+$ and $\pi^0$ mesons are attached to the incoming and outgoing baryon lines, respectively, whereas in panels (\subref{fig:fey_dig_2e})--(\subref{fig:fey_dig_2h}), the attachment order is reversed.}
\label{fig:fey_dig_2}
\end{figure}
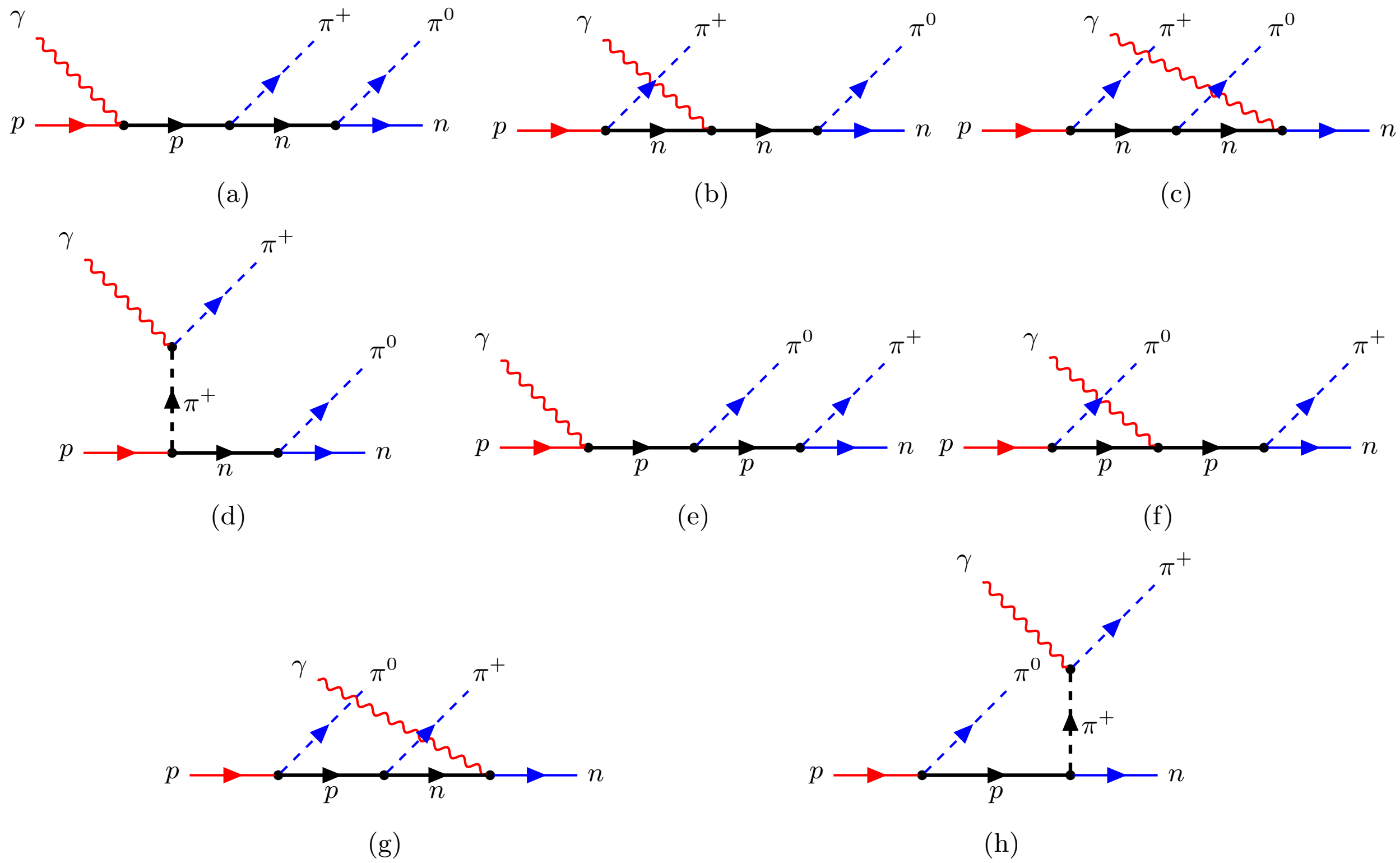

The calculation of the invariant amplitudes for this process involves two classes of diagrams, distinguished by which pion couples to the incoming and outgoing baryon lines, as shown in Fig.~\ref{fig:fey_dig_2}. The amplitudes corresponding to the first class of diagrams, shown in Fig.~\ref{fig:fey_dig_2}(\subref{fig:fey_dig_2a})–(\subref{fig:fey_dig_2d}), are given by 
\begin{align}
    \mathcal{M}^{(s_1)} =~& \frac{-i e \ g_{\pi^0 n n} \ g_{\pi^+ p n}}{[(p_2+q_2)^2-m^2][(p_1+k)^2-m^2]} \ \bar{u} (p_2)  \ (\slashed{p}_2 + \slashed{q}_2 - m) \nonumber \\ 
    & \times  (\slashed{p}_1 + \slashed{k} + m)  \left[ \slashed{\epsilon}  - \frac{\kappa_p}{4m}  (\slashed{\epsilon} \slashed{k} -\slashed{k} \slashed{\epsilon} )  \right] u (p_1) \ , \\
    \mathcal{M}^{(u_{11})} =~& \frac{-i e  \ g_{\pi^0 nn} \ g_{\pi^+ pn}}{[(p_2+q_2)^2-m^2][(p_1-q_1)^2-m^2]} \ \bar{u} (p_2)  \ (\slashed{p}_2 + \slashed{q}_2 - m) \nonumber \\
    &\times  \frac{\kappa_n}{4m}  (\slashed{\epsilon} \slashed{k} - \slashed{k} \slashed{\epsilon} )   (\slashed{p}_1 -\slashed{q}_1 - m) \ u (p_1) \ , \\
    \mathcal{M}^{(u_{12})}=~& \frac{-i e \ g_{\pi^0 n n} \ g_{\pi^+ p n}}{[(p_2-k)^2-m^2][(p_1-q_1)^2-m^2]} \ \bar{u} (p_2)  \left[ - \frac{\kappa_n}{4m}  (\slashed{\epsilon} \slashed{k} - \slashed{k} \slashed{\epsilon})  \right] \nonumber \\
    & \times (\slashed{p}_2 - \slashed{k} + m)   (\slashed{p}_1 -\slashed{q}_1 -m) \ u (p_1) \ , \\
    \mathcal{M}^{(t_1)} =~& \frac{-ie \ g_{\pi^0 n n} \ g_{\pi^+ p n}}{[(p_2+q_2)^2-m^2][(k-q_1)^2-m_{\pi^+}^2]} \ \bar{u} (p_2)  \ (\slashed{p}_2 + \slashed{q}_2 - m)  [(2q_1 -k) \cdot \epsilon] u (p_1)   \ ,
\end{align}
where $\kappa_n$ denotes the anomalous magnetic moment of the neutron. As in the previous case, the amplitudes $\mathcal{M}^{(s_1)}$ and $\mathcal{M}^{(t_1)}$ are not gauge invariant individually. However, gauge invariance is restored when their electric terms are combined, as demonstrated by
\begin{align}
   \left[ \mathcal{M}^{(s_1)}_{\rm el} + \mathcal{M}^{(t_1)}_{\rm el} \right]_{\epsilon^\mu\to k^\mu}  =& -i e \ g_{\pi^0 n n} \ g_{\pi^+ p n} \ \bar{u} (p_2)\biggl\{ \frac{(\slashed{p}_2 + \slashed{q}_2 - m)(\slashed{p}_1 + \slashed{k} + m) \slashed{k}}{[(p_2+q_2)^2-m^2][(p_1+k)^2-m^2]} \nonumber \\
    &+ \frac{(2q_1 -k) \cdot k\, (\slashed{p}_2 +\slashed{q}_2 - m) }{[(k - q_1)^2-m_{\pi^+}^2][(p_2 + q_2)^2-m^2]} \biggr\} u (p_1) \nonumber \\
    ~=~& -i e \ g_{\pi^0 n n} \ g_{\pi^+ p n} \ \bar{u} (p_2)\biggl\{ \frac{(\slashed{p}_2 + \slashed{q}_2 - m)}{(p_2+q_2)^2-m^2} \biggl[ \frac{(\slashed{p}_1 + \slashed{k} + m) \slashed{k}}{(p_1+k)^2-m^2} \nonumber \\
    & + \frac{(2q_1 -k) \cdot k }{[(k - q_1)^2-m_{\pi^+}^2]}\biggr] \biggr\} u (p_1) \nonumber \\
    ~=~& 0 \ .
\end{align}
For the second class of diagrams shown in Fig.~\ref{fig:fey_dig_2}(\subref{fig:fey_dig_2e}), (\subref{fig:fey_dig_2f}), (\subref{fig:fey_dig_2g}), and (\subref{fig:fey_dig_2h}), the amplitudes read
\begin{align}
    \mathcal{M}^{(s_2)} ~=~& \frac{-i e \ g_{\pi^+ p n} \ g_{\pi^0 p p}}{[(p_2+q_1)^2-m^2][(p_1+k)^2-m^2]} \ \bar{u} (p_2)  \ (\slashed{p}_2 + \slashed{q}_1 - m) \nonumber \\ 
    ~& \times  (\slashed{p}_1 + \slashed{k} + m)  \left[ \slashed{\epsilon}  - \frac{\kappa_p}{4m}  (\slashed{\epsilon} \slashed{k} -\slashed{k} \slashed{\epsilon} )  \right] u (p_1) \ ,  \\
    \mathcal{M}^{(u_{21})} ~=~& \frac{-i e \ g_{\pi^+ p n} \ g_{\pi^0 p p}}{[(p_2+q_1)^2-m^2][(p_1-q_2)^2-m^2]} \ \bar{u} (p_2)  \ (\slashed{p}_2 + \slashed{q}_1 - m) \nonumber \\
    ~&\times \left[ \slashed{\epsilon}  + \frac{\kappa_p }{4m}  (\slashed{\epsilon} \slashed{k} - \slashed{k} \slashed{\epsilon} )  \right] (\slashed{p}_1 -\slashed{q}_2 - m) \ u (p_1) \ , \\
    \mathcal{M}^{(u_{22})} 
    ~=~& \frac{-i e \ g_{\pi^+ p n} \ g_{\pi^0 p p}}{[(p_2-k)^2-m^2][(p_1-q_2)^2-m^2]} \ \bar{u} (p_2)  \left[- \frac{\kappa_n}{4m}  (\slashed{\epsilon} \slashed{k} - \slashed{k} \slashed{\epsilon})  \right] \nonumber \\
    ~& \times (\slashed{p}_2 - \slashed{k} + m)   (\slashed{p}_1 -\slashed{q}_2 -m) \ u (p_1) \ , \\
    \mathcal{M}^{(t_2)} ~=~&\frac{-i e \ g_{\pi^+ p n} \ g_{\pi^0 p p}}{[(k - q_1)^2-m_{\pi^+}^2][(p_1 - q_2)^2-m^2]} \ \bar{u} (p_2)  \   [(2q_1 -k) \cdot \epsilon](\slashed{p}_1 -\slashed{q}_2 - m)  u (p_1) \ .
\end{align}
Here, the amplitudes $\mathcal{M}^{(s_2)}$, $\mathcal{M}^{(u_{21})}$, and $\mathcal{M}^{(t_2)}$ are likewise not gauge invariant individually. Gauge invariance is restored when their electric terms are combined, as can be demonstrated by
\begin{align}
    \MoveEqLeft 
    \left[ \mathcal{M}^{(s_2)}_{\rm el} + \mathcal{M}^{(u_{21})}_{\rm el} + \mathcal{M}^{(t_2)}_{\rm el} \right]_{\epsilon^\mu\to k^\mu} \nonumber\\
    =& -i  e \ g_{\pi^+ p n} \ g_{\pi^0 p p} \ \bar{u} (p_2)\biggl\{ \frac{(\slashed{p}_2 + \slashed{q}_1 - m)(\slashed{p}_1 + \slashed{k} + m) \slashed{k}}{[(p_2+q_1)^2-m^2][(p_1+k)^2-m^2]} + \frac{(\slashed{p}_2 + \slashed{q}_1 - m)\slashed{k}(\slashed{p}_1 - \slashed{q}_2 - m) }{[(p_2+q_1)^2-m^2][(p_1-q_2)^2-m^2]} \nonumber \\
    &+ \frac{(2q_1 -k) \cdot k\, (\slashed{p}_1 -\slashed{q}_2 - m) }{[(k - q_1)^2-m_{\pi^+}^2][(p_1 - q_2)^2-m^2]} \biggr\} u (p_1) \nonumber \\
    =& -i  e \ g_{\pi^+ p n} \ g_{\pi^0 p p} \ \bar{u} (p_2)\biggl\{ \frac{(\slashed{p}_2 + \slashed{q}_1 - m)}{(p_2+q_1)^2-m^2} \biggl[ \frac{(\slashed{p}_1 + \slashed{k} + m) \slashed{k}}{(p_1+k)^2-m^2} + \frac{\slashed{k}(\slashed{p}_1 - \slashed{q}_2 - m) }{(p_1-q_2)^2-m^2} \biggr] \nonumber \\
    &+ \frac{(2q_1 -k) \cdot k(\slashed{p}_1 -\slashed{q}_2 - m) }{[(k - q_1)^2-m_{\pi^+}^2][(p_1 - q_2)^2-m^2]} \biggr\} u (p_1) \nonumber \\
    =& ~0 \ .
\end{align}
The Dirac amplitudes $A_i$ obtained from the decomposition of the $\gamma p \to \pi^+ \pi^0 n$ production amplitude read
\begin{align}
    A^{(\mathbf{B})}_1 =& -\frac{\kappa_p}{2m} \left[ \left(\frac{k^2+2p_1\cdot k}{2}\right) \big(C^{(s_1)}+C^{(s_2)}\big) - \big(q_1\cdot k - 2 q \cdot Q - q_1 \cdot q_2 + 2p_1 \cdot q\big)C^{(u_{21})} \right]
    \nonumber \\
    & -\frac{\kappa_n}{2m} \left[ \left(\frac{k^2-2p_2\cdot k}{2}\right) \big(C^{(u_{12})}+C^{(u_{22})}\big) - \big(q_2\cdot k + 2 q \cdot Q - q_1 \cdot q_2 - 2p_1 \cdot q\big)C^{(u_{11})} \right] \ , \\
    A^{(\mathbf{B})}_2 = & \frac{1}{m}\left\{\frac{q\cdot k}{P \cdot k} \left[\kappa_p\big(2C^{(s_1)}-2C^{(s_2)} + C^{(u_{21})}\big) 
    -\kappa_n \ C^{(u_{11})}\right] +\kappa_n\,C^{(u_{11})} +\kappa_p\,C^{(u_{21})}\right\} \ , \\
    A^{(\mathbf{B})}_3 = & \frac{1}{m\,P\cdot k}\Big[-2\kappa_p(Q\cdot k+P\cdot k)\big(C^{(s_1)}-C^{(s_2)}\big) 
    +(Q\cdot k-P\cdot k)\big(\kappa_n C^{(u_{11})}-\kappa_p C^{(u_{21})}\big)\Big] \ , \\
    A^{(\mathbf{B})}_5 =& \big(C^{(t_1)}-C^{(t_2)}\big) -\kappa_n\,C^{(u_{11})} -(1+\kappa_p)\,C^{(u_{21})} \ , \\
    A^{(\mathbf{B})}_6 =& (1+\kappa_p)\big(2C^{(s_1)}-2C^{(s_2)} - C^{(u_{21})}\big) +\big(C^{(t_1)}-C^{(t_2)}\big) 
    +\kappa_n\,C^{(u_{11})}  \ , \\
    A^{(\mathbf{B})}_7 =& (1+\kappa_p)\big(C^{(s_1)}-C^{(s_2)}-C^{(u_{21})}\big) 
    +\kappa_n\big(C^{(u_{11})}+C^{(u_{12})}-C^{(u_{22})}\big) \ , \\
    A^{(\mathbf{B})}_8 =& \frac{1}{q_2\cdot k\ P\cdot k}\Big[\big(C^{(s_1)}-C^{(s_2)}-C^{(u_{21})}\big)(k^2+2p_1\cdot k-2q_2\cdot k) \nonumber\\
    &-(C^{(t_1)} - C^{(t_2)})(k^2-4q\cdot k) \Big] \ , \\
    A^{(\mathbf{B})}_9 =& \frac{1}{q_2\cdot k\ P\cdot k}\Big[-\big(C^{(s_1)}-C^{(s_2)}-C^{(u_{21})}\big)(k^2+2p_1\cdot k) \nonumber\\
    &+ (C^{(t_1)} - C^{(t_2)})(k^2- 4 Q\cdot k) \Big] \ , \\
    A^{(\mathbf{B})}_{10} =&-\frac{1}{m}\Big[\kappa_p\big(C^{(s_1)}-C^{(s_2)}-C^{(u_{21})}\big) 
    +\kappa_n\big(C^{(u_{11})}+C^{(u_{12})}-C^{(u_{22})}\big)\Big] \ , \\
    A^{(\mathbf{B})}_{11} =&\frac{1}{m}\Big[\kappa_n\big(C^{(u_{12})}-C^{(u_{22})}\big) -\kappa_p\big(C^{(s_1)}-C^{(s_2)}\big)\Big] \ , \\
    A^{(\mathbf{B})}_{12} =&\frac{1}{m}\Big[\kappa_n\,C^{(u_{11})}+\kappa_p\,C^{(u_{21})}\Big] \ , \\
    A^{(\mathbf{B})}_{13} =&-\big(C^{(t_1)}-C^{(t_2)}\big) \ , \\
    A^{(\mathbf{B})}_{16} =& \frac{C^{(t_1)}-C^{(t_2)}}{P\cdot k} \ , \\
    A^{(\mathbf{B})}_{4} =& A^{(\mathbf{B})}_{14} = A^{(\mathbf{B})}_{15} = 0 \ ,
\end{align}
where $C^{(s_i)}$, $C^{(t_i)}$ and $C^{(u_{ij})}$, with $i, j \in \{1, 2\}$, denote the products of the coupling constants and propagators associated with the corresponding diagrams, defined as
\begin{align}
    C^{(s_1)} =~& \frac{-i e \ g_{\pi^0 n n} \ g_{\pi^+ p n}}{[(p_2+q_2)^2-m^2][(p_1+k)^2-m^2]}  \ , \\
    C^{(s_2)} =~& \frac{-i e \ g_{\pi^+ p n} \ g_{\pi^0 p p}}{[(p_2+q_1)^2-m^2][(p_1+k)^2-m^2]}  \ ,   \\
    C^{(t_1)} =~& \frac{-i e \ g_{\pi^0 n n} \ g_{\pi^+ p n}}{[(p_2+q_2)^2-m^2][(k-q_1)^2-m_{\pi^+}^2]}    \ , \\
    C^{(t_2)} =~&\frac{-i  e \ g_{\pi^+ p n} \ g_{\pi^0 p p}}{[(k - q_1)^2-m_{\pi^+}^2][(p_1 - q_2)^2-m^2]}  \ ,  \\
    C^{(u_{11})} =~& \frac{-i  e  \ g_{\pi^0 nn} \ g_{\pi^+ pn}}{[(p_2+q_2)^2-m^2][(p_1-q_1)^2-m^2]}  \ , \\
    C^{(u_{12})}=~& \frac{-i e \ g_{\pi^0 n n} \ g_{\pi^+ p n}}{[(p_2-k)^2-m^2][(p_1-q_1)^2-m^2]} \ , \\
    C^{(u_{21})} =~& \frac{-i e \ g_{\pi^+ p n} \ g_{\pi^0 p p}}{[(p_2+q_1)^2-m^2][(p_1-q_2)^2-m^2]} \ , \\
    C^{(u_{22})} =~& \frac{-i  e \ g_{\pi^+ p n} \ g_{\pi^0 p p}}{[(p_2-k)^2-m^2][(p_1-q_2)^2-m^2]}  \ .
\end{align}

\subsection{Hadronic Form Factor}
Since hadrons are composite particles, all hadronic vertices in the diagrams shown in Figs.~\ref{fig:fey_dig_1} and \ref{fig:fey_dig_2} should be supplemented by form factors. A more rigorous treatment of form factors in double-meson photo- and electroproduction that maintains gauge invariance can be found in, e.g., Ref.~\cite{Haberzettl:2018hcl}. Here, we account for them in a phenomenological manner, using the commonly adopted dipole form \cite{Haberzettl:1998aqi}
\begin{align}
\label{eq:dipole_ff}
F(q) = \frac{\Lambda^{4}}{\Lambda^{4}+\left(q^{2}-m^{2}\right)^{2}} \ ,
\end{align}
where $q$ and $m$ are the four-momentum and the mass of the internal particle, respectively, and $\Lambda$ is a cutoff parameter. The form factor is normalized to unity, $F(q)=1$, when the internal particle is on its mass shell, i.e., $q^{2}=m^{2}$. In contrast to single-meson photoproduction, however, each diagram in double-meson photoproduction contains two internal lines. Consequently, the form factor of the $i$-th diagram is the product
\begin{align}
\label{eq:ff_diagram}
F_{i}=F\!\left(q_{i,1}\right) F\!\left(q_{i,2}\right) \ .
\end{align}
It is important to emphasize that a single hadronic vertex may be attached to two off-shell lines simultaneously, e.g., in the diagrams shown in Figs.~\ref{fig:fey_dig_1}(a) and \ref{fig:fey_dig_2}(a). In such cases, the form factor of that vertex should depend on both momenta \cite{Haberzettl:1998aqi}, $F\!\left(q_{i,1},q_{i,2}\right)$, and the total form factor cannot in general be written as the product of two single-line factors given in Eq.~\eqref{eq:ff_diagram}. As a first approximation, however, we retain the product form of Eq.~\eqref{eq:ff_diagram} in the present study and leave a detailed analysis of this issue for future work.

Note, however, that introducing form factors in this way violates gauge invariance, since diagrams whose sum is gauge invariant are multiplied by different factors, and the cancellation among them is destroyed. To restore gauge invariance, we multiply every diagram of a given group by the common form factor \cite{Davidson:2001rk}
\begin{align}
\label{eq:ff_common}
    F_{c}=1-\prod_{i}\left(1-F_{i}\right) \ ,
\end{align}
where the product runs over those diagrams that are not individually gauge invariant but whose sum is, and $F_{i}$ is the form factor of the $i$-th diagram defined in Eq.~\eqref{eq:ff_diagram}. Since a single overall factor multiplies the whole group, it factors out of the sum over the diagrams, and gauge invariance is preserved. Equation~\eqref{eq:ff_common} also has the required limits, i.e., $F_{c}\to 1$ whenever any one of the $F_{i}$ approaches unity, so that the on-shell behavior of the amplitude is left unchanged, while $F_{c}\to 0$ when all internal particles are far off shell.\footnote{Note that Ref.~\cite{Davidson:2001rk} used a similar form factor for single-pion photoproduction, $\hat{F}(s,u,t) = F_1(s) + F_1(u) + F_3(t) - F_1(s)F_1(u) - F_1(s)F_3(t) - F_1(u)F_3(t) + F_1(s)F_1(u)F_3(t),$ which is a special case of Eq.~\eqref{eq:ff_common}, where the product runs over the $s$-, $u$-, and $t$-channel Born diagrams and the same function $F_1$ is used in the $s$ and $u$ channels. This choice is intended to preserve the crossing symmetry of the amplitude in single-pion photoproduction.} For simplicity, in this exploratory study we use a single cutoff $\Lambda$ for all internal lines and examine its effect on the observables.

\subsection{Numerical Results}
\label{sec:numerical}
We now evaluate the total cross sections and invariant-mass distributions for $\gamma p\to\pi^0\pi^0p$ and $\gamma p\to\pi^+\pi^0n$ from the Dirac amplitudes derived in Secs.~\ref{sec:res_pi0_pi0} and \ref{sec:res_pip_pi0}, inserted into the cross-section formula of Sec.~\ref{subsec:The_cross_section}, and compare the results with the data measured with the Crystal Ball/TAPS setup at MAMI~\cite{Zehr:2012tj}. Since only nucleon Born terms are included, we restrict the comparison to the near-threshold region, $E_\gamma^{\rm lab}\le 400$~MeV, i.e., from the thresholds at $E_\gamma^{\rm lab}=308.8$~MeV ($\pi^0\pi^0p$) and 316.4~MeV ($\pi^+\pi^0n$) up to a total c.m. energy of $W\approx 1.28$~GeV. Note that this range already covers the $\Delta(1232)$ peak in the $\gamma p$ system, so that contributions involving the $\Delta(1232)$ are expected to become important with increasing energy. The production amplitudes of both processes depend on only a few parameters, namely the $\pi NN$ coupling constant and the anomalous magnetic moments of the nucleons,
\begin{equation}
    g_{\pi^0pp}=13.3, \quad \kappa_p = 1.792847, \quad \kappa_n = -1.913043,
\end{equation}
together with the isospin relations between the coupling constants,
\begin{equation}
    g_{\pi^0nn} = -g_{\pi^0pp}, \quad g_{\pi^+pn}=\sqrt{2}\,g_{\pi^0pp}.
\end{equation}
None of these parameters is fitted to the double-pion data. The only free parameter is the cutoff $\Lambda$ of the hadronic form factor in Eq.~\eqref{eq:dipole_ff}, which we vary between 700 and 1000~MeV in order to assess the sensitivity of the results to the off-shell behavior of the Born amplitudes.

\begin{figure}[ht]
\includegraphics[width=\columnwidth]{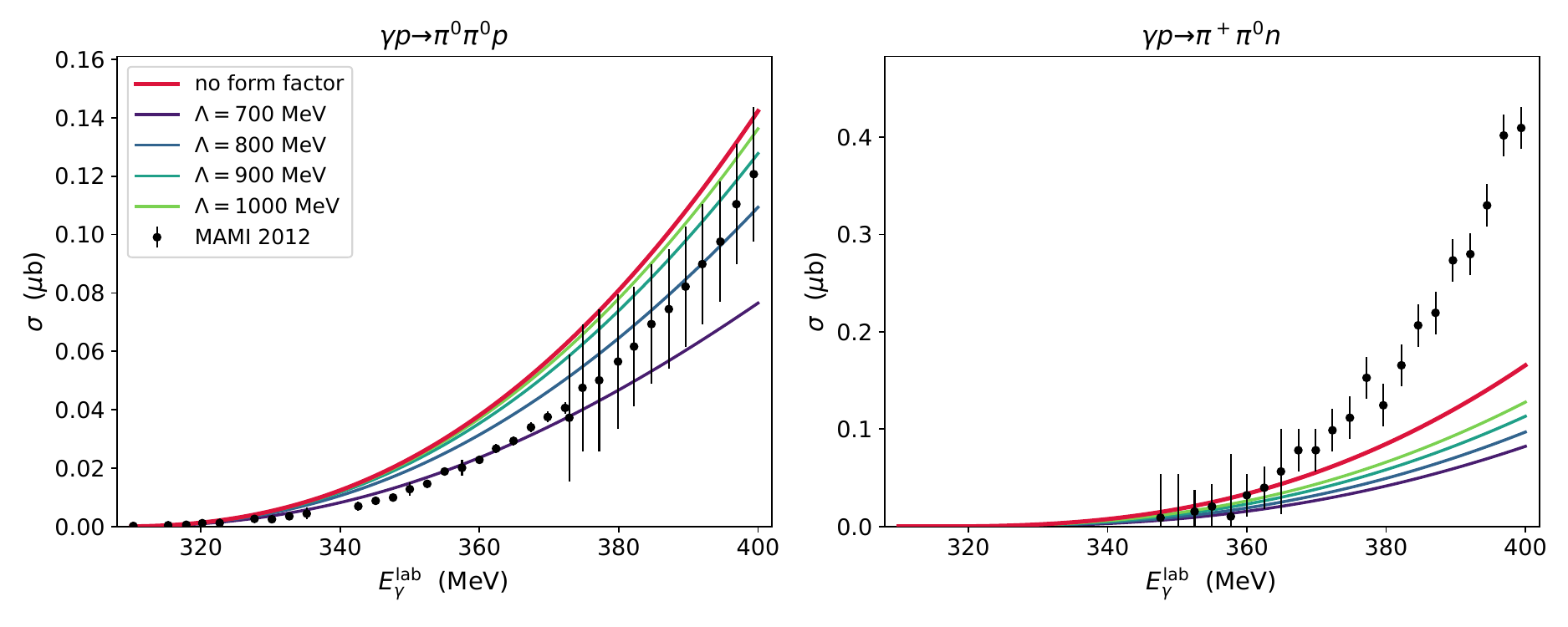}
\caption{\label{fig:total_cs} Total cross sections for $\gamma p\to\pi^0\pi^0p$ (left) and $\gamma p\to\pi^+\pi^0n$ (right) as functions of the photon laboratory energy $E_\gamma^{\rm lab}$, calculated from the nucleon Born terms of Figs.~\ref{fig:fey_dig_1} and \ref{fig:fey_dig_2}. The thick red curves are obtained without hadronic form factors, whereas the thin curves are obtained with the gauge-invariant common form factor of Eq.~\eqref{eq:ff_common} and cutoff masses $\Lambda=700$, 800, 900, and 1000~MeV. The data are taken from Ref.~\cite{Zehr:2012tj}.}
\end{figure}

The total cross sections are shown in Fig.~\ref{fig:total_cs}. Since only a very limited set of diagrams is included, a quantitative description of the data cannot be expected; the purpose of the comparison is rather to identify the extent to which the Born terms account for the measured cross sections and to locate the missing strength. For comparison, the effective-Lagrangian model of Ref.~\cite{Fix:2005if}, which includes about twenty classes of resonant and nonresonant diagrams, among them the $\Delta(1232)$, $\rho$, and $\sigma$ contributions omitted here, describes the data up to $E_\gamma^{\rm lab}\approx 1.5$~GeV, but still underestimates the $\pi^0\pi^0p$ cross section near threshold. The two channels behave quite differently in our calculation, and we discuss them in turn.

For $\gamma p\to\pi^0\pi^0p$ (left panel of Fig.~\ref{fig:total_cs}), the Born terms without form factors overestimate the data over the whole energy range, by a factor of roughly 1.5--2 between $E_\gamma^{\rm lab}\approx 340$ and 370~MeV. The form factor reduces the cross section, and its effect grows with energy, since the internal nucleon lines are increasingly far off shell. With $\Lambda=700$~MeV, the calculation agrees with the data from threshold up to $E_\gamma^{\rm lab}\approx 370$~MeV. Above this energy, however, the data rise more steeply than the $\Lambda=700$~MeV curve and are, within their larger uncertainties, better matched by the curves with $\Lambda=800$--900~MeV. Consequently, no single cutoff describes the whole energy range, which indicates that the missing contributions increase with energy. The most natural candidates are the $\Delta(1232)$ contributions, both as an intermediate state in the $s$- and $u$-channel diagrams and in the $p\pi^0$ subsystem, whose importance grows as the $\Delta(1232)$ region is approached, and the isoscalar $S$-wave $\pi\pi$ final-state interaction, which in the covariant ChPT analysis of Ref.~\cite{Kang:2024fsf} is required to describe these data above $E_\gamma^{\rm lab}\approx 340$~MeV.

The agreement obtained with $\Lambda=700$~MeV should, however, not be interpreted as evidence that the $\pi^0\pi^0p$ channel near threshold is dominated by nucleon Born terms. In ChPT, where the $\pi N$ interaction is constrained by chiral symmetry, the tree-level contributions to this channel are strongly suppressed near threshold, and the threshold cross section is dominated by pion loops, i.e., by charged-pion pairs that rescatter into $\pi^0\pi^0$ \cite{Bernard:1994ds,Bernard:1996ft}. This prediction is supported by the MAMI data~\cite{Kotulla:2003cx,Zehr:2012tj}. Correspondingly, the tree-level amplitude of Ref.~\cite{Kang:2024fsf}, which includes nonresonant terms up to second chiral order together with the $\Delta(1232)$, $N(1440)$, and $\rho$ contributions, lies well below the data. In contrast, the pseudoscalar $\pi NN$ coupling used here is not constrained by chiral symmetry, and for two-pion production it is not equivalent to the pseudovector coupling, since the transformation relating the two couplings generates additional interactions of second order in the pion field, which contribute when two pions are attached to the same nucleon line. As a result, the size of the Born contribution to the $\pi^0\pi^0p$ channel depends strongly on the choice of coupling \cite{Hirata:2002tp,Fix:2005if}. The comparison in the left panel of Fig.~\ref{fig:total_cs} therefore demonstrates that the formalism yields sensible results with the expected threshold behavior, but the physical content of the near-threshold $\pi^0\pi^0p$ amplitude can only be assessed once the coupling scheme, the corresponding contact terms, and the $\pi\pi$ rescattering are treated consistently.

For $\gamma p\to\pi^+\pi^0n$ (right panel of Fig.~\ref{fig:total_cs}), the situation is reversed. The Born terms are compatible with the data only up to $E_\gamma^{\rm lab}\approx 365$~MeV, where the experimental uncertainties are large. Beyond this region they rise much more slowly than the data, and at $E_\gamma^{\rm lab}=400$~MeV they account for only about 40\% of the measured cross section. Since in our calculation the form factor reduces the cross section further, to about 20\% of the data for $\Lambda=700$~MeV, this discrepancy cannot be removed by adjusting $\Lambda$; it signals the absence of important production mechanisms. Unlike the $\pi^0\pi^0$ final state, the $\pi^+\pi^0n$ final state can be produced through the $\Delta$ Kroll-Ruderman and $\Delta$ pion-pole mechanisms, e.g., $\gamma p\to\pi^+\Delta^0\to\pi^+\pi^0n$, and through $\rho^+$ production \cite{GomezTejedor:1995pe,Fix:2005if}. Our finding is consistent with that of Ref.~\cite{Kang:2024fsf}, where the nonresonant chiral amplitude alone falls far below the $\pi^+\pi^0n$ data, the $\Delta(1232)$ contributions are found to be indispensable below $E_\gamma^{\rm lab}\approx 440$~MeV, and the $\rho$ meson further enhances the cross section near threshold. It is therefore the charged-pion channel, rather than the neutral one, that exposes most clearly the incompleteness of a pure nucleon-Born description.

\begin{figure}[ht]
\includegraphics[width=\columnwidth]{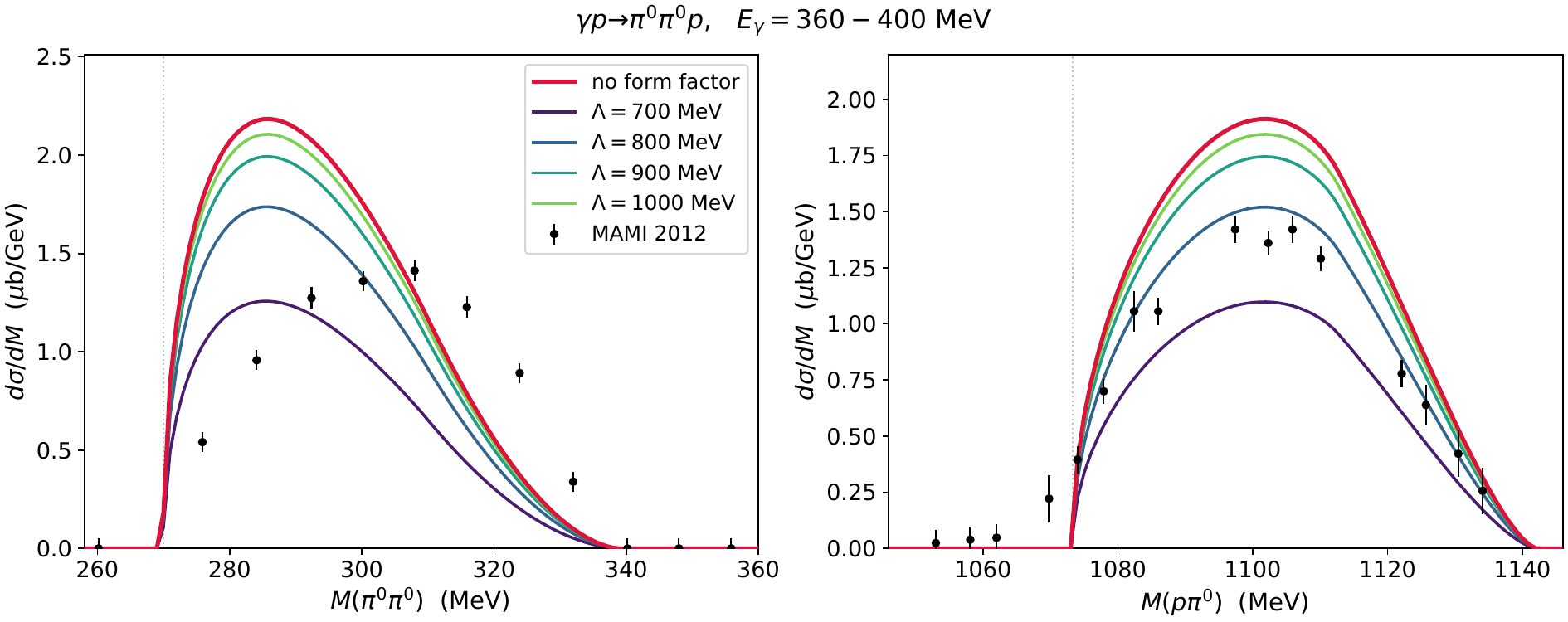}
\caption{\label{fig:d_cs} Invariant-mass distributions of the $\pi^0\pi^0$ (left) and $p\pi^0$ (right) systems for $\gamma p\to\pi^0\pi^0p$, averaged over the photon energy bin $E_\gamma^{\rm lab}=360$--$400$~MeV. The vertical dotted lines indicate the $\pi^0\pi^0$ and $p\pi^0$ thresholds. The curves are as in Fig.~\ref{fig:total_cs}, and the data are taken from Ref.~\cite{Zehr:2012tj}.}
\end{figure}

The total cross sections integrate over the five-dimensional phase space and can therefore be described by quite different combinations of mechanisms. A more stringent test is provided by the invariant-mass distributions, which are shown in Fig.~\ref{fig:d_cs} for $\gamma p\to\pi^0\pi^0p$ in the energy bin $E_\gamma^{\rm lab}=360$--$400$~MeV. In this bin, the $\pi^0\pi^0$ invariant mass extends from $2m_{\pi^0}=270$~MeV up to at most 339~MeV, whereas the $p\pi^0$ invariant mass extends from 1073~MeV up to at most 1142~MeV.

The $p\pi^0$ distribution (right panel of Fig.~\ref{fig:d_cs}) is well reproduced in shape by all curves, which differ essentially only in their normalization; the curve with $\Lambda=800$~MeV comes closest to the data also in magnitude. This agreement should not be overrated, because in this energy bin the $p\pi^0$ spectrum is still close to the shape of three-body phase space, which peaks at about 1100~MeV, as do both the data and the calculation. The $\pi^0\pi^0$ distribution (left panel of Fig.~\ref{fig:d_cs}) is considerably more discriminating. Whereas three-body phase space would peak at about 297~MeV, the Born terms shift the strength towards the $2\pi^0$ threshold and peak at about 285~MeV, while the data peak at about 305~MeV. Consequently, the calculated spectra rise much more steeply above threshold than the data and fall off earlier. At $M(\pi^0\pi^0)\approx 276$~MeV all curves exceed the data by about a factor of two or more, whereas between 316 and 324~MeV even the curve without form factor only reaches the data, and the curve with $\Lambda=700$~MeV, which describes the total cross section at lower energies, falls well below them. Since the form factor essentially rescales the distribution, this discrepancy in shape cannot be removed by varying $\Lambda$. It reveals that the missing dynamics resides mainly in the $\pi\pi$ subsystem. The natural candidate is the strong attraction in the isoscalar $S$-wave $\pi\pi$ interaction associated with the $\sigma$ [$f_0(500)$] meson. Since the corresponding phase shift rises from threshold, the rescattering enhancement is expected to grow with the $\pi\pi$ invariant mass and thus to shift the strength towards higher masses, in the direction required by the data. This interpretation is consistent with the importance of $\pi\pi$ rescattering and pion loops found in the chiral analyses of this channel \cite{Bernard:1994ds,Roca:2002vd,Kang:2024fsf}.

These results show that, although the Born terms provide a useful benchmark, a realistic description of double-pion photoproduction near threshold requires at least the $\Delta(1232)$ contributions, the $\rho$ meson in the charged channel, and the $\pi\pi$ final-state interaction in the isoscalar $S$ wave. Within the present formalism, all of these mechanisms enter simply as additional contributions to the Dirac amplitudes $A_i$; for the $\Delta(1232)$, the use of consistent spin-3/2 couplings is advisable \cite{Pascalutsa:2000kd,Vrancx:2011qv,Kristiano:2017qjq,Mart:2019jtb}. At the same time, the example illustrates a central advantage of double-meson production emphasized in Sec.~\ref{sec:intro}: a calculation that gives the right size of the cross section can still fail to describe how the two mesons share the available energy, as seen in the $\pi^0\pi^0p$ invariant-mass spectrum.

\section{Summary and Conclusion}
\label{sec:summary_and_conclusion}

We have developed a general formalism for the electromagnetic production of two mesons on the nucleon, $\gamma_{\rm v}N\to\Phi_1\Phi_2B$, that extends to three-body final states the approach introduced in the 1960s by Berends \textit{et al.} for single-meson production. Starting from a general Lorentz-covariant expansion of the hadronic current and imposing current conservation, we have expressed the production amplitude in terms of 16 gauge-invariant matrices $M_i$ and the corresponding Lorentz-invariant Dirac amplitudes $A_i$. Four of these matrices vanish for real photons. Reducing the Dirac structures to Pauli operators, we have obtained 16 CGLN-like Pauli amplitudes $\mathcal{F}_j$, related to the $A_i$ by a $16\times 16$ kinematical matrix, and have expressed the five-fold differential cross section of virtual photoproduction in terms of the transverse, longitudinal, and transverse-transverse and transverse-longitudinal interference response functions, in complete analogy to single-meson electroproduction. Despite the linear dependence among some of these matrices, they follow immediately from the covariant decomposition of the current. This makes it possible to read off the amplitudes $A_i$ from the Feynman diagrams and yields more compact, symmetric expressions for the observables. Since the Dirac amplitudes are the only model-dependent input, the formalism can be combined with any reaction model, from tree-level effective Lagrangians to unitarized and coupled-channel approaches. The derivation of the remaining polarization observables within this framework is left for future work.

As a first application, we have calculated the total cross sections for $\gamma p\to\pi^0\pi^0p$ and $\gamma p\to\pi^+\pi^0n$ as functions of the photon laboratory energy, as well as the $\pi^0\pi^0$ and $p\pi^0$ invariant-mass distributions for $\gamma p\to\pi^0\pi^0p$, using only nucleon Born terms with pseudoscalar $\pi N$ coupling. This example shows how the Dirac amplitudes are read off directly from the Feynman diagrams. It also demonstrates explicitly that gauge invariance, which is violated by the individual diagrams, is restored in the total amplitude, and that it can be preserved after the introduction of hadronic form factors by multiplying each gauge-invariant group of diagrams by a common form factor.

The comparison with the MAMI data in the near-threshold region leads to the following conclusions. (i) For $\gamma p\to\pi^0\pi^0p$, the Born terms overestimate the total cross section; a soft form factor with $\Lambda=700$~MeV reproduces the data up to $E_\gamma^{\rm lab}\approx 370$~MeV, but no single cutoff describes the whole energy range up to 400~MeV. Since tree-level contributions to this channel are suppressed in chiral perturbation theory and the Born strength depends strongly on the $\pi NN$ coupling scheme, this agreement should be regarded as effective rather than as evidence for Born-term dominance. (ii) For $\gamma p\to\pi^+\pi^0n$, the Born terms account for only about 40\% of the measured cross section at $E_\gamma^{\rm lab}=400$~MeV, and form factors reduce it further; the missing strength must therefore come from additional mechanisms, most notably the $\Delta(1232)$ Kroll-Ruderman and pion-pole terms and $\rho$-meson production. (iii) The $p\pi^0$ invariant-mass distribution is reproduced in shape, whereas the calculated $\pi^0\pi^0$ distribution peaks about 20~MeV too close to threshold, independently of the cutoff. This points to the isoscalar $S$-wave $\pi\pi$ interaction, i.e., the $\sigma$ [$f_0(500)$] channel, as an essential missing ingredient, and illustrates that invariant-mass distributions can rule out descriptions that reproduce the integrated cross section in the same energy range.

These findings define a clear program for future work. The natural next steps are the inclusion of the $\Delta(1232)$ with consistent spin-3/2 couplings \cite{Pascalutsa:2000kd,Vrancx:2011qv}, the $\rho$ and $\sigma$ mesons, the pseudovector $\pi NN$ coupling together with the associated contact terms, and the $\pi\pi$ and $\pi N$ final-state interactions, followed by an extension to the second resonance region, where the $N(1440)$, $N(1520)$, and $N(1535)$ resonances become important. Within the present formalism, each of these mechanisms simply adds a contribution to the Dirac amplitudes $A_i$, while the expressions for the observables remain unchanged. Because the formalism is formulated for virtual photons from the outset, the same amplitudes can be used directly to analyze double-pion electroproduction, for which a wealth of data from CLAS and CLAS12 is now available \cite{CLAS:2023mfc,CLAS:2026qhw,CLAS:2026ywm,Mokeev:2026plj}. We therefore expect the present framework to provide a convenient and versatile basis for theoretical and phenomenological studies of double-meson photo- and electroproduction, including coupled-channel analyses aimed at the extraction of nucleon resonance properties.

\begin{acknowledgments}
R. M. P. and T. M. were supported by the PUTI Q1 Grant from Universitas Indonesia under Contract No. PKS-206/UN2.RST/HKP.05.00/2025. S. C. was supported by the Young Scientist Training (YST) Program at the Asia Pacific Center for Theoretical Physics (APCTP) through the Science and Technology Promotion Fund and Lottery Fund of the Korean Government and also by the Korean Local Governments--Gyeongsangbuk-do Province and Pohang City.
\end{acknowledgments}

\appendix

\section{Relations between the amplitudes $A_i$ and $B_i$}
\label{app:relation_Ai_Bi}
The relations between the Lorentz-invariant amplitudes $A_i$ and $B_i$ are
\begin{align} \label{eq:amp_Aj}
    A_1 &= B_1 \ , \nonumber \\
    A_2 &= \frac{-k^2 \ B_1-\left(p_1- q_2\right) \cdot k \ B_2-\left(p_2+q_2\right)\cdot k~B_3-2q\cdot k~B_4}{ P \cdot k~q_2 \cdot k  } \ , \nonumber \\
    A_3 &= \frac{k^2 \ B_1+ p_1 \cdot k \ B_2+p_2\cdot k \ B_3+2Q \cdot k \ B_4}{ P \cdot k \ q_2 \cdot k } \ , \nonumber \\
    A_4 &= B_7+B_8\ , \nonumber \\
    A_5 &= B_7-B_8+B_9+B_{10}\ , \nonumber \\
    A_6 &= B_9-B_{10}\ , \nonumber \\
    A_7 &= B_{11} \ ,  \nonumber\\
    A_8 &= \frac{-k^2 \ B_{11}-\left(p_1- q_2\right) \cdot k \ B_{12}-\left(p_2+q_2\right)\cdot k \ B_{13}-2q\cdot k \ B_{14}}{ P \cdot k \ q_2 \cdot k }\ ,  \\
    A_9 &= \frac{k^2 \ B_{11}+p_1 \cdot k \ B_{12}+p_2\cdot k \ B_{13} +2Q \cdot k \ B_{14}}{ P \cdot k \ q_2 \cdot k }\ ,  \nonumber\\
    A_{10} &= B_{17}+B_{18} \ , \nonumber \\
    A_{11} &= B_{17}-B_{18}+B_{19}+B_{20}\ , \nonumber \\
    A_{12} &= B_{19} - B_{20}\ , \nonumber \\
    A_{13} &= B_8-B_7\ , \nonumber \\
    A_{14} &= \frac{2B_1-B_2+B_3 }{2P \cdot k}\ , \nonumber \\
    A_{15} &= B_{18}-B_{17}\ , \nonumber \\
    A_{16} &= \frac{2B_{11}-B_{12}+B_{13} }{2P \cdot k} \ . \nonumber
\end{align}

\section{Pauli Amplitudes $\mathcal{F}_j$}
\label{app:Pauli_amplitudes}
The Pauli amplitudes $\mathcal{F}_j$ are related to the Dirac amplitudes $A_i$ through
\begin{align}
\mathcal{F}_1={}& (\tilde k_0-\check k_0)A_1
-\beta b_{QP} A_2
-\beta(b_q P_0-b_P q_0)A_3
+\Big(\tilde P_0\check k_0+P_0\tilde k_0-\tilde P_0 k_0-b_P\Big)A_4
+\Big(\tilde Q_0\check k_0+Q_0\tilde k_0-\tilde Q_0 k_0
\nonumber\\
&-b_Q\Big)A_5+\Big(\tilde q_0\check k_0+q_0\tilde k_0-\tilde q_0 k_0-b_q\Big)A_6
-\Big\{2(q_0\tilde k_0-\tilde q_0 k_0+\tilde q_0\check k_0)+\check q_0 k_0-\check k_0 q_0-\check q_0\tilde k_0-b_q\Big\}A_7
\nonumber\\
&+(\beta q_0-\tilde q_0-\check q_0)\,b_{QP}\,A_8
+(\beta q_0-\tilde q_0-\check q_0)\,b_{qP}\,A_9+\Big\{2\beta b_{qP}-\tilde P_0(b_q-q_0\check k_0)+\check q_0(b_P-P_0\tilde k_0)\Big\}A_{10}\nonumber\\
&+\Big\{2\beta b_{qQ}-\tilde Q_0(b_q-q_0\check k_0)+\check q_0(b_Q-Q_0\tilde k_0)\Big\}A_{11}
+\Big\{q_0(\tilde q_0\check k_0-\tilde k_0\check q_0)+(\check q_0-\tilde q_0)b_q\Big\}A_{12}
+\frac{1}{2}(\tilde k_0\check k_0-k^2)A_{13}
\nonumber\\
&-\beta b_{kP} A_{14}
+\frac{1}{2}\Big\{2\beta b_{qk}-\tilde k_0(b_q-q_0\check k_0)+\check q_0(k^2-k_0\tilde k_0)\Big\}A_{15}+(\beta q_0-\tilde q_0-\check q_0)\,b_{kP}\,A_{16} ,\\[2ex]
\mathcal{F}_2={}& \mathrm{q\,k}\sin\vartheta\left(-A_7+\frac{b_{QP}}{W_1}A_8+\frac{b_{qP}}{W_1}A_9-\tilde P_0 A_{10}-\tilde Q_0 A_{11}-\tilde q_0 A_{12}-\frac{1}{2}\tilde k_0 A_{15}+\frac{b_{kP}}{W_1}A_{16}\right),\\[2ex]
\mathcal{F}_3={}&\frac{2\mathrm{Q\,k}}{W_2}\bigg(A_1
-\frac{b_{QP}}{W_1}A_2
-\frac{b_{qP}}{W_1}A_3
-\tilde P_0 A_4
-\tilde Q_0 A_5
-\tilde q_0 A_6
+2\tilde q_0 A_7
-\frac{2b_{Pq}}{W_1}A_{10}
-\frac{2b_{Qq}}{W_1}A_{11}
\nonumber\\
&-\frac{1}{2}\tilde k_0 A_{13}
-\frac{b_{kP}}{W_1}A_{14}
-\frac{b_{kq}}{W_1}A_{15}
\bigg)
+\mathrm{k}\,\left(\mathrm{q}\cos\vartheta+\frac{2\mathrm{Q}}{W_2}q_0\right)\bigg(-\,A_7
+\frac{\, b_{QP}}{W_1}A_8
+\frac{\, b_{qP}}{W_1}A_9
\nonumber\\
&-\,\tilde P_0A_{10}
-\,\tilde Q_0A_{11}
-\,\tilde q_0 A_{12}
-\frac{1}{2}\,\tilde k_0A_{15}
+\frac{\, b_{kP}}{W_1}A_{16}\bigg),\\[2ex]
\mathcal{F}_4={}& \frac{2\mathrm{Q\,q}}{W_2}\sin\vartheta\bigg\{-(\tilde k_0-k_0)A_7+b_{QP} A_8+b_{qP} A_9+(P_0\tilde k_0-b_P)A_{10}+(Q_0\tilde k_0-b_Q)A_{11}\nonumber\\
&+(q_0\tilde k_0-b_q)A_{12}+\frac{1}{2}(k_0\tilde k_0-k^2)A_{15}+b_{kP} A_{16}\bigg\},\\[2ex]
\mathcal{F}_5={}& \mathrm{k}\bigg[\alpha A_1+\frac{b_P}{W_1}A_4+\frac{b_Q}{W_1}A_5+\frac{b_q}{W_1}A_6+\big\{2(\tilde q_0-\alpha q_0)+\alpha \check q_0\big\}A_7-\frac{\check q_0 b_P}{W_1}A_{10}-\frac{\check q_0 b_Q}{W_1}A_{11}\nonumber\\
&-\frac{\check q_0 b_q}{W_1}A_{12}+\frac{k^2}{2W_1}A_{13}-\frac{\check q_0 k^2}{2W_1}A_{15}\bigg],\\[2ex]
\mathcal{F}_6={}& -\mathrm{q}\sin\vartheta\left(\tilde k_0 A_7+b_P A_{10}+b_Q A_{11}+b_q A_{12}+\frac{1}{2}k^2 A_{15}\right),\\[2ex]
\mathcal{F}_7={}& \frac{2\mathrm{Q}}{W_2}\bigg(\tilde k_0 A_1-b_P A_4-b_Q A_5-b_q A_6+2b_q A_7-\frac{1}{2}k^2 A_{13}\bigg)\nonumber\\
&-\left(\mathrm{q}\cos\vartheta+\frac{2\mathrm{Q}}{W_2}q_0\right)\bigg(\tilde k_0 A_7+ b_P A_{10}+ b_Q A_{11}+ b_q A_{12}+\frac{1}{2} k^2 A_{15}\bigg),\\[2ex]
\mathcal{F}_8={}& \frac{2\mathrm{Q\,q\,k}}{W_2}\sin\vartheta\bigg(-\alpha A_7+\frac{b_P}{W_1}A_{10}+\frac{b_Q}{W_1}A_{11}+\frac{b_q}{W_1}A_{12}+\frac{k^2}{2W_1}A_{15}\bigg),\\[2ex]
\mathcal{F}_9={}& \mathrm{q}\sin\vartheta\bigg[-\beta b_P A_3-\Big\{\tilde k_0+\alpha(\check k_0-k_0)\Big\}A_6+\Big\{\tilde k_0+2\alpha(\check k_0-k_0)\Big\}A_7+(\beta q_0-\tilde q_0-\check q_0)\,b_P\,A_9
\nonumber\\
&+(2\beta-1)b_P A_{10}+(2\beta-1)b_Q A_{11}-\Big\{k_0\tilde q_0-\tilde k_0\check q_0+\alpha q_0(\check k_0-k_0)\Big\}A_{12}+\frac{1}{2}(2\beta-1)k^2 A_{15}\bigg],\\[2ex]
\mathcal{F}_{10}={}& \mathrm{q^2\,k}\sin^2\vartheta\left(\frac{b_P}{W_1}A_9+\alpha A_{12}\right),\\[2ex]
\mathcal{F}_{11}={}& \mathrm{q\,k}\sin\vartheta\Bigg\{\frac{2\mathrm{Q}}{W_2}\left(-\frac{b_P}{W_1}A_3+\alpha A_6-2\alpha A_7+\frac{2b_P}{W_1}A_{10}+\frac{2b_Q}{W_1}A_{11}+\frac{k^2}{W_1}A_{15}\right)\nonumber\\
&+\left(\mathrm{q}\cos\vartheta+\frac{2\mathrm{Q}}{W_2}q_0\right)\bigg(\frac{ b_P}{W_1}A_9+\alpha \, A_{12}\bigg)\Bigg\},\\[2ex]
\mathcal{F}_{12}={}& \frac{2\mathrm{Q\,q^2}}{W_2}\sin^2\vartheta\left(b_P A_9-\tilde k_0 A_{12}\right),\\[2ex]
\mathcal{F}_{13}={}& \cot\vartheta \,\mathcal{F}_9+\mathrm{Q}\,\Bigg[\frac{2\tilde k_0}{W_2}A_1-\beta c_1 A_2-\beta b_q A_3
+\bigg\{\tilde k_0+\alpha(\check k_0-k_0)-\frac{2b_P}{W_2}\bigg\}A_4
-\bigg\{\tilde k_0+\alpha(\check k_0-k_0)+\frac{2b_Q}{W_2}\bigg\}A_5
\nonumber\\
&-\frac{2b_q}{W_2}A_6
+\frac{2}{W_2}(2b_q-q_0\tilde k_0)A_7
+(\beta q_0-\tilde q_0-\check q_0)\,c_1\,A_8
+(\beta q_0-\tilde q_0-\check q_0)\,b_q\,A_9
+\bigg\{k_0\tilde q_0-\tilde k_0\check q_0\nonumber\\
&+(\check k_0-k_0)(2\tilde q_0-\alpha q_0)+b_q-\frac{2q_0 b_P}{W_2}\bigg\}A_{10}
-\bigg\{k_0\tilde q_0-\tilde k_0\check q_0+(\check k_0-k_0)(2\tilde q_0-\alpha q_0)+b_q+\frac{2b_Q q_0}{W_2}\bigg\}A_{11}
\nonumber\\
&-\frac{2q_0 b_q}{W_2}A_{12}-\frac{k^2}{W_2}A_{13}-\beta k^2 A_{14}-\frac{q_0 k^2}{W_2}A_{15}+(\beta q_0-\tilde q_0-\check q_0)\,k^2\,A_{16}\Bigg],\\[2ex]
\mathcal{F}_{14}={}& \cot\vartheta\,\mathcal{F}_{10}+\mathrm{Q\,q\,k}\sin\vartheta\,\bigg(\frac{c_1}{W_1}A_8+\frac{b_q}{W_1}A_9-\alpha A_{10}+\alpha A_{11}+\frac{k^2}{W_1}A_{16}\bigg),\\[2ex]
\mathcal{F}_{15}={}& -\cot^2\vartheta\,\mathcal{F}_{10}+\cot\vartheta\,\left(\mathcal{F}_{11}+\mathcal{F}_{14}\right)+\frac{2\mathrm{Q^2\,k}}{W_2}\bigg(-\frac{c_1}{W_1}A_2-\frac{b_q}{W_1}A_3-\alpha A_4+\alpha A_5+\frac{q_0 c_1}{W_1}A_8+\frac{q_0 b_q}{W_1}A_9\nonumber\\
&-(2\tilde q_0-\alpha q_0)A_{10}+(2\tilde q_0-\alpha q_0)A_{11}-\frac{k^2}{W_1}A_{14}+\frac{q_0 k^2}{W_1}A_{16}\bigg),\\[2ex]
\mathcal{F}_{16}={}& \cot\vartheta\, \mathcal{F}_{12}+\frac{2\mathrm{Q^2\,q}}{W_2}\sin\vartheta\,\Big(c_1 A_8+b_q A_9+\tilde k_0 A_{10}-\tilde k_0 A_{11}+k^2 A_{16}\Big).
\end{align}
Note that we have defined $\mathrm{P}$, $\mathrm{Q}$ and $\mathrm{k}$ as magnitudes of the vectors $\bm{P}$, $\bm{Q}$ and $\bm{k}$ respectively. Also, we introduced the following notation:
\begin{align}
    W_1 ={}& E_1 + m_1 \ , & W_2 ={}& E_2 + m_2 \ , & c_1 ={}& (s-m_1^2)/2\ ,\nonumber\\
    \alpha ={}& 1+\frac{k_0}{W_1}\ , & \beta ={}& 1-\frac{2\bm{Q}\cdot\bm{k}}{W_1W_2},
\end{align}
and some generalized symbol:
\begin{align}
    b_X \equiv{}& X\cdot  k\ , & b_{XY}\equiv{}& b_XY_0 - b_YX_0 \ ,\nonumber\\
    \tilde{X}\equiv{}& X_0+\frac{\bm{X}\cdot\bm{k}}{W_1} \ , &\check{X}\equiv{}& X_0 +\frac{2\bm{X}\cdot\bm{Q}}{W_2}\ ,
\end{align}
where $X,Y$ are any four-vector from $P,\, Q,\, q$ and $k$. Note also that since $b_k = k^2$, we use $k^2$ instead of the notation $b_k$ in our formulations.

\section{Components of the hadronic tensor}
\label{app:had_tensor}

The components of the hadronic tensor $H^{\mu\nu}$ relevant for the response functions and the differential cross section discussed in Sec.~\ref{subsec:The_cross_section} are
\begin{align}
    \MoveEqLeft {\sum_{\lambda_N,\lambda_B} \frac{|J_x|^2+|J_y|^2}{4\mathcal{N}^2}} \nonumber\\
    =&\;|\mathcal{F}_5|^2+|\mathcal{F}_8|^2 +\mathcal{F}_{5,10}+\mathcal{F}_{5,15}+\mathcal{F}_{6,16}-\mathcal{F}_{7,12}-\mathcal{F}_{8,11}+\mathcal{F}_{8,14}+\tfrac{1}{2}\big(|\mathcal{F}_6|^2+|\mathcal{F}_7|^2+|\mathcal{F}_9|^2+|\mathcal{F}_{10}|^2
    \nonumber\\    &+|\mathcal{F}_{11}|^2+|\mathcal{F}_{12}|^2+|\mathcal{F}_{13}|^2+|\mathcal{F}_{14}|^2+|\mathcal{F}_{15}|^2+|\mathcal{F}_{16}|^2\big)+\big(2\mathcal{F}_{5,7}-\mathcal{F}_{5,12}+\mathcal{F}_{6,8}+\mathcal{F}_{7,10}+\mathcal{F}_{7,15}
    \nonumber\\
    &+\mathcal{F}_{8,9}-\mathcal{F}_{10,12}-\mathcal{F}_{14,16}\big)\cos\theta+\tfrac{1}{2}\big(|\mathcal{F}_7|^2-|\mathcal{F}_{11}|^2-|\mathcal{F}_{13}|^2-|\mathcal{F}_{14}|^2-2|\mathcal{F}_{15}|^2-|\mathcal{F}_{16}|^2 -2\mathcal{F}_{5,15}
    \nonumber\\
    &-2\mathcal{F}_{6,16}+2\mathcal{F}_{8,11}-2\mathcal{F}_{8,14}\big)\cos^2\theta - \big(\mathcal{F}_{7,15}-\mathcal{F}_{14,16}\big)\cos^3\theta + \tfrac{1}{2} \ | \mathcal{F}_{15}|^2\cos^4\theta
    - \big(2\mathcal{F}_{5,6}+\mathcal{F}_{5,16}
    \nonumber\\
    &+\mathcal{F}_{6,10}+\mathcal{F}_{6,15}-\mathcal{F}_{7,8}-\mathcal{F}_{8,13}+\mathcal{F}_{11,12}+\mathcal{F}_{15,16}\big)\cos\varphi\sin\theta +\tfrac{1}{2}\big(|\mathcal{F}_6|^2-|\mathcal{F}_{9}|^2-2|\mathcal{F}_{10}|^2-|\mathcal{F}_{11}|^2
    \nonumber \\
    &-|\mathcal{F}_{12}|^2-|\mathcal{F}_{14}|^2 -2\mathcal{F}_{5,10} +2\mathcal{F}_{7,12}
    +2\mathcal{F}_{8,11}-2\mathcal{F}_{8,14}\big)\cos^2\varphi\sin^2\theta +\big(\mathcal{F}_{6,10}+\mathcal{F}_{11,12}\big)\cos^3\varphi\sin^3\theta
    \nonumber\\
    & + \tfrac{1}{2} \ |\mathcal{F}_{10}|^2\cos^4\varphi\sin^4\theta
     + \big(\mathcal{F}_{5,11}+\mathcal{F}_{5,14}-\mathcal{F}_{6,7}+\mathcal{F}_{6,12}-\mathcal{F}_{7,16}+\mathcal{F}_{9,13}+\mathcal{F}_{10,11}+\mathcal{F}_{10,14}\nonumber\\
    &+\mathcal{F}_{11,15}+\mathcal{F}_{12,16}+\mathcal{F}_{14,15}\big)\cos\varphi\sin\theta\cos\theta - \big(\mathcal{F}_{6,11} +\mathcal{F}_{6,14}+\mathcal{F}_{7,10}-\mathcal{F}_{10,12}+\mathcal{F}_{11,16}\nonumber\\
    &+\mathcal{F}_{12,15}\big)\cos^2\varphi\sin^2\theta\cos\theta - \big(\mathcal{F}_{10,11}+ \mathcal{F}_{10,14}\big)\cos^3\varphi\sin^3\theta\cos\theta  + \big(\mathcal{F}_{6,15}+\mathcal{F}_{7,11}+\mathcal{F}_{7,14}-\mathcal{F}_{10,16}\nonumber \\
    &-\mathcal{F}_{12,14}+\mathcal{F}_{15,16}\big)\cos\varphi\sin\theta\cos^2\theta+\tfrac{1}{2}\big(|\mathcal{F}_{11}|^2+|\mathcal{F}_{14}|^2+2\mathcal{F}_{10,15}+2\mathcal{F}_{11,14}\big)\cos^2\varphi\sin^2\theta\cos^2\theta  \nonumber \\
    &- \big(\mathcal{F}_{11,15} + \mathcal{F}_{14,15}\big)\cos\varphi\sin\theta\cos^3\theta,\\[2ex]
\label{eq:jx2-jy2}
    \MoveEqLeft\sum_{\lambda_N,\lambda_B} \frac{|J_x|^2-|J_y|^2}{4\mathcal{N}^2} \nonumber\\
    =&\;\frac{1}{2}
  \Big[-2\big(\mathcal{F}_{5,12} + \mathcal{F}_{6,8} - \mathcal{F}_{8,9}\big)\cos2\varphi\cos\theta-2\big(\mathcal{F}_{5,16} + \mathcal{F}_{7,8} - \mathcal{F}_{8,13}\big)\cos\varphi\sin\theta  \nonumber\\
 &+\Big\{-|\mathcal{F}_7|^2+ |\mathcal{F}_{13}|^2+|\mathcal{F}_{14}|^2+|\mathcal{F}_{15}|^2 + |\mathcal{F}_{16}|^2+ 2\mathcal{F}_{5,15}+ 2\mathcal{F}_{6,16}+ 2\big(\mathcal{F}_{7,15}- \mathcal{F}_{14,16}\big)\cos\theta\nonumber\\
 &-|\mathcal{F}_{15}|^2\cos^2\theta- 2\big(\mathcal{F}_{6,15}+\mathcal{F}_{15,16})\cos\varphi\sin\theta-|\mathcal{F}_{14}|^2\cos^2\varphi\sin^2\theta+\mathcal{F}_{14,15}\cos\varphi\sin2\theta\nonumber\\
 &+2\big(\mathcal{F}_{8,11} + \mathcal{F}_{8,14}\big)\sin^2\varphi\Big\}\sin^2\theta+\mathcal{F}_{8,10}\sin\varphi\sin2\varphi\sin2\theta\nonumber\\
 &+\Big\{\mathcal{F}_{5,11} + \mathcal{F}_{5,14}- \mathcal{F}_{6,7}+ \mathcal{F}_{6,12} - \mathcal{F}_{7,16} + \mathcal{F}_{9,13} + \mathcal{F}_{10,14} +\mathcal{F}_{11,15} + \mathcal{F}_{12,16} \nonumber\\
 & + \big(\mathcal{F}_{7,11} + \mathcal{F}_{7,14}- \mathcal{F}_{10,16}-\mathcal{F}_{12,14}\big)\cos\theta - \mathcal{F}_{11,15}\cos^2\theta- \big(\mathcal{F}_{6,11}+ \mathcal{F}_{6,14}+ \mathcal{F}_{11,16}  \nonumber\\
 & + \mathcal{F}_{12,15}\big)\cos\varphi\sin\theta-\mathcal{F}_{10,14}\cos^2\varphi\sin^2\theta + \big(\mathcal{F}_{10,15}+\mathcal{F}_{11,14})\cos\varphi\sin\theta\cos\theta \Big\}\cos\varphi\sin2\theta\nonumber\\
 &+\Big\{-|\mathcal{F}_6|^2 + |\mathcal{F}_9|^2+ |\mathcal{F}_{10}|^2+ |\mathcal{F}_{11}|^2 + |\mathcal{F}_{12}|^2+ 2\mathcal{F}_{5,10}- 2\mathcal{F}_{7,12}+ 2\big(\mathcal{F}_{7,10}-\mathcal{F}_{10,12}\big)\cos\theta\nonumber\\
 & -|\mathcal{F}_{11}|^2\cos^2\theta- 2\big(\mathcal{F}_{6,10}+\mathcal{F}_{11,12}\big)\cos\varphi\sin\theta -|\mathcal{F}_{10}|^2\cos^2\varphi\sin^2\theta\nonumber\\
 & + \mathcal{F}_{10,11}\cos\varphi\sin2\theta\Big\}\big(\cos^2\theta\cos^2\varphi-\sin^2\varphi\big)\Big]\cos2\phi \nonumber\\
  &+\frac{1}{2}\Big[\big(\mathcal{F}_{5,12} + \mathcal{F}_{6,8} - \mathcal{F}_{8,9}\big)\sin2\varphi\big(1+\cos^2\theta\big)+\big(\mathcal{F}_{5,16} + \mathcal{F}_{7,8} - \mathcal{F}_{8,13}\big)\sin\varphi\sin2\theta\nonumber\\
  & -2\Big\{\mathcal{F}_{5,11} + \mathcal{F}_{5,14}- \mathcal{F}_{6,7}+ \mathcal{F}_{6,12} - \mathcal{F}_{7,16} + \mathcal{F}_{9,13} + \mathcal{F}_{10,14} +\mathcal{F}_{11,15} + \mathcal{F}_{12,16} \nonumber\\
 & + \big(\mathcal{F}_{7,11} + \mathcal{F}_{7,14}- \mathcal{F}_{10,16}-\mathcal{F}_{12,14}\big)\cos\theta - \mathcal{F}_{11,15}\cos^2\theta- \big(\mathcal{F}_{6,11}+ \mathcal{F}_{6,14}+ \mathcal{F}_{11,16}  \nonumber\\
 & + \mathcal{F}_{12,15}\big)\cos\varphi\sin\theta-\mathcal{F}_{10,14}\cos^2\varphi\sin^2\theta + \big(\mathcal{F}_{10,15}+\mathcal{F}_{11,14})\cos\varphi\sin\theta\cos\theta\Big\}\sin\varphi\sin\theta\nonumber\\
  &-\Big\{-|\mathcal{F}_6|^2 + |\mathcal{F}_9|^2+ |\mathcal{F}_{10}|^2+ |\mathcal{F}_{11}|^2 + |\mathcal{F}_{12}|^2+ 2\mathcal{F}_{5,10}- 2\mathcal{F}_{7,12}+ 2\big(\mathcal{F}_{7,10}-\mathcal{F}_{10,12}\big)\cos\theta\nonumber\\
 & -|\mathcal{F}_{11}|^2\cos^2\theta- 2\big(\mathcal{F}_{6,10}+\mathcal{F}_{11,12}\big)\cos\varphi\sin\theta -|\mathcal{F}_{10}|^2\cos^2\varphi\sin^2\theta\nonumber\\
 & + \mathcal{F}_{10,11}\cos\varphi\sin2\theta\Big\}\sin2\varphi\cos\theta + 2\mathcal{F}_{8,10}\big(\cos^2\theta\cos^2\varphi-\sin^2\varphi\big)\sin\varphi\sin\theta\nonumber\\
 &+2\mathcal{F}_{8,15}\sin\varphi\sin^3\theta+\big(\mathcal{F}_{8,11} + \mathcal{F}_{8,14}\big)\sin2\varphi\sin^2\theta\cos\theta\Big]\sin2\phi ,\\[2ex]
\MoveEqLeft\sum_{\lambda_N,\lambda_B} \frac{|J_z|^2}{2\mathcal{N}^2}\nonumber\\
=&\; \frac{k_0^2}{|\bm{k}|^2}\Big[
|\mathcal{F}_1|^2 + \sin^2\varphi\,|\mathcal{F}_2|^2 + |\mathcal{F}_3|^2 + |\mathcal{F}_4|^2 - 2\mathcal{F}_{2,4}\,\cos\theta + \big(\cos^2\varphi\,|\mathcal{F}_2|^2 - |\mathcal{F}_3|^2\big)\cos^2\theta \nonumber\\
&- 2\cos\varphi\,\mathcal{F}_{3,4}\,\sin\theta + 2\cos\varphi\,\mathcal{F}_{2,3}\,\sin\theta\cos\theta\Big],\\[2ex]
\MoveEqLeft\sum_{\lambda_N,\lambda_B} \frac{\text{Re}(J_x^*J_z)}{2\mathcal{N}^2} \nonumber\\
=&\; \frac{k_0}{|\bm{k}|}\Bigg\{\bigg[
\big(\mathcal{F}_{1,8} - \mathcal{F}_{4,5}\big)\cos\varphi
+ \big(\mathcal{F}_{1,9}+\mathcal{F}_{2,5}+\mathcal{F}_{2,10}-\mathcal{F}_{4,7}+\mathcal{F}_{4,12}\big)\cos\varphi\cos\theta
+ \big(\mathcal{F}_{2,7}-\mathcal{F}_{2,12}\nonumber\\
&-\mathcal{F}_{4,10}\big)\cos\varphi\cos^2\theta
+ \big(\mathcal{F}_{1,13}+\mathcal{F}_{2,8}\sin^2\varphi+\mathcal{F}_{2,14}+\mathcal{F}_{3,5}+\mathcal{F}_{4,6}+\mathcal{F}_{4,16}\big)\sin\theta
\nonumber\\
&-\Big\{\big(\mathcal{F}_{2,6}+\mathcal{F}_{3,12}+\mathcal{F}_{4,11}\big)\cos^2\varphi-\mathcal{F}_{3,7}+\mathcal{F}_{2,16}+\mathcal{F}_{4,14}\Big\}\sin\theta\cos\theta
\nonumber\\
&+ \big(\mathcal{F}_{2,11}+\mathcal{F}_{3,10}\big)\sin\theta\cos^2\theta\cos^2\varphi
-\big(\mathcal{F}_{3,6}+\mathcal{F}_{3,16}+\mathcal{F}_{4,15}\big)\sin^2\theta\cos\varphi
\nonumber\\
&+ \Big\{\big(\mathcal{F}_{2,15}+\mathcal{F}_{3,11}+\mathcal{F}_{3,14}\big)\cos\varphi-\mathcal{F}_{2,10}\cos^3\varphi\Big\}\sin^2\theta\cos\theta
+ \big(\mathcal{F}_{3,15}-\cos^2\varphi\,\mathcal{F}_{2,14}\big)\sin^3\theta
\bigg]\cos\phi
\nonumber\\
&+ \bigg[
-\big(\mathcal{F}_{1,9}+\mathcal{F}_{2,5}+\mathcal{F}_{2,10}-\mathcal{F}_{4,7}+\mathcal{F}_{4,12}\big)\sin\varphi
- \big(\mathcal{F}_{1,8}+\mathcal{F}_{2,7}-\mathcal{F}_{2,12}-\mathcal{F}_{4,5}-\mathcal{F}_{4,10}\big)\sin\varphi\cos\theta
\nonumber\\
&+ \big(\mathcal{F}_{2,6}+\mathcal{F}_{3,12}+\mathcal{F}_{4,11}\big)\sin\theta\cos\varphi\sin\varphi
+ \big(\mathcal{F}_{2,8}-\mathcal{F}_{2,11}-\mathcal{F}_{3,10}\big)\sin\theta\cos\theta\cos\varphi\sin\varphi
\nonumber\\
&+ \big(\mathcal{F}_{3,8}-\mathcal{F}_{3,11}+\cos^2\varphi\,\mathcal{F}_{2,10}\big)\sin^2\theta\sin\varphi
\bigg]\sin\phi\Bigg\} ,
\label{eq:re_jxs_jz}
\end{align}
where
\begin{align}
    \mathcal{F}_{i,j}= \text{Re}(\mathcal{F}_i\mathcal{F}_j^*) \ .
\end{align}


\bibliography{ref}

\end{document}